\documentclass[pra,twocolumn,amsmath,amssymb,superscriptaddress]{revtex4-2}

\usepackage{graphicx}
\usepackage{dcolumn}
\usepackage{bm}
\usepackage{color}
\usepackage{appendix}

\usepackage{graphicx}
\usepackage{amsthm,amssymb}
\usepackage{array}
\usepackage{hyperref}
\usepackage{xcolor}
\usepackage{mathtools,upgreek}
\usepackage{bm,times,enumitem}
\hypersetup{colorlinks=true, urlcolor={black}, linkcolor={blue!80!black}, citecolor={blue!80!black}}

\newcommand{\Tr}[1]{\mathrm{Tr}\{#1\}}
\newcommand{\sgn}[1]{\mathrm{sgn}(#1)}

 \newcommand{\ket}[1]{|{#1}\rangle}
 \newcommand{\bra}[1]{\langle{#1}|}

\begin{document}

\title{Emulation of quantum measurements with mixtures of coherent states}
\author{A.~Mikhalychev}
\affiliation{B. I. Stepanov Institute of Physics, NAS of Belarus, Nezavisimosti ave. 68, 220072 Minsk, Belarus}
\author{Y. S. Teo}
\affiliation{Department of Physics and Astronomy, Seoul National University, 08826 Seoul, South Korea}
\author{H. Jeong}
\affiliation{Department of Physics and Astronomy, Seoul National University, 08826 Seoul, South Korea}
\author{A.~Stefanov}
\affiliation{Institute of Applied Physics, University of Bern, Sidlerstrasse 5, CH-3012 Bern, Switzerland}
\author{D.~Mogilevtsev}
\affiliation{B. I. Stepanov Institute of Physics, NAS of Belarus, Nezavisimosti ave. 68, 220072 Minsk, Belarus}
\email{d.mogilevtsev@ifanbel.bas-net.by}
\date{February 2022}

\begin{abstract}
We propose a methodology to emulate quantum phenomena arising from any non-classical quantum state using only a finite set of mixtures of coherent states. This allows us to successfully reproduce well-known quantum effects using resources that can be much more feasibly generated in the laboratory. We present a simple procedure to experimentally carry out quantum-state emulation with coherent states, illustrate it emulating multi-photon NOON states with few phase-averaged coherent states, and demonstrate its capabilities in observing fundamental quantum-mechanical effects, such as the Hong-Ou-Mandel effect, violating Bell inequalities and witnessing quantum non-classicality. 
\end{abstract}

\maketitle

\section{Introduction}

Understanding the extent to which classical elements can be used to reveal nontrivial quantum effects in experiments can offer a deeper perspective on the interplay between classical and quantum resources. On the one hand, {there are cornerstone results obtained  with essentially non-classical states}, such as antibunching of photons and suppression of the field amplitude noise below the classical level~\cite{PhysRevLett.59.2044,RevModPhys.68.127}, exceeding the standard quantum limit in measurement precision~\cite{metro}, violating Bell inequalities~\cite{PhysicsPhysiqueFizika.1.195}, and the exhibition of non-classicality signatures~\cite{PhysRev.130.2529,PhysRevLett.10.277,PhysRev.140.B676}. On the other hand, interesting new research on how these quantum effects can still be observed using classical resources have emerged. For example, recently it was shown that Bell inequalities violation and other quantum-like signatures may be brought to light by ``classical entanglement'', that is, by local classical correlations of different degrees of freedom~\cite{Qian:15,Goldin:10,Karimi1172,Toppel:15,spre,Khrennikov2020}.   

In this work we investigate the potential of emulating experiments over a quantum state using a set of {"classical" probe states, i.e., those possessing a non-negative Glauber-Sudarshan P functions~\cite{PhysRevLett.10.277,PhysRev.131.2766}, examples of which include mixtures of coherent states}. The underlying principle behind emulating an arbitrary state with the density matrix $\rho$ is the linear-algebraic fact that $\rho$ is expressible as a linear combination of a set of non-orthogonal basis states, where some of the coefficients in such a linear combination can be negative. This approach underlies the so-called ``data pattern'' method developed for quantum tomography \cite{PhysRevLett.105.010402,Mogilevtsev_2013,BayesianPRA2015,Motka_2017,reut2017}, and allows one to avoid calibrating the measurement setup by fitting a response from an unknown state to the responses from other known probe states \cite{pub.1039448817}. We suggest the scheme that works in somewhat opposite way: we fit the state with ``probes'' aiming to produce the same measured response as the true quantum state. In a typical scenario in quantum mechanics, the observer has access to the complete quantum state $\rho$, with which the probabilities of all possible experimental outcomes can be predicted.  The essence of our quantum-state emulation is to employ a classical preparation procedure that samples the basis states we want in order to reproduce {measurement results obtained with this $\rho$: either of a particular measurement or of an arbitrary one (under certain reasonable constraints)}. The respective coefficients that go with the basis states are incorporated through post-processing. Our emulation scheme may be understood as an \emph{a priori} assignment of classical information that specifies the components of $\rho$ during state preparation, and is otherwise absent in a non-emulation (usual) scenario. {Noticeably, fitting of a particular measurement does not require high fidelity of the representation with the true quantum state. On the other hand, high-fidelity representation implies accurate fitting of the measurement results for \textit{any} observable with limited eigenvalues. We show how to achieve such representation of exquisitely quantum multi-photon NOON states by using just few phase-averaged coherent states}. 

\begin{figure}[t]
	\includegraphics[width=\linewidth]{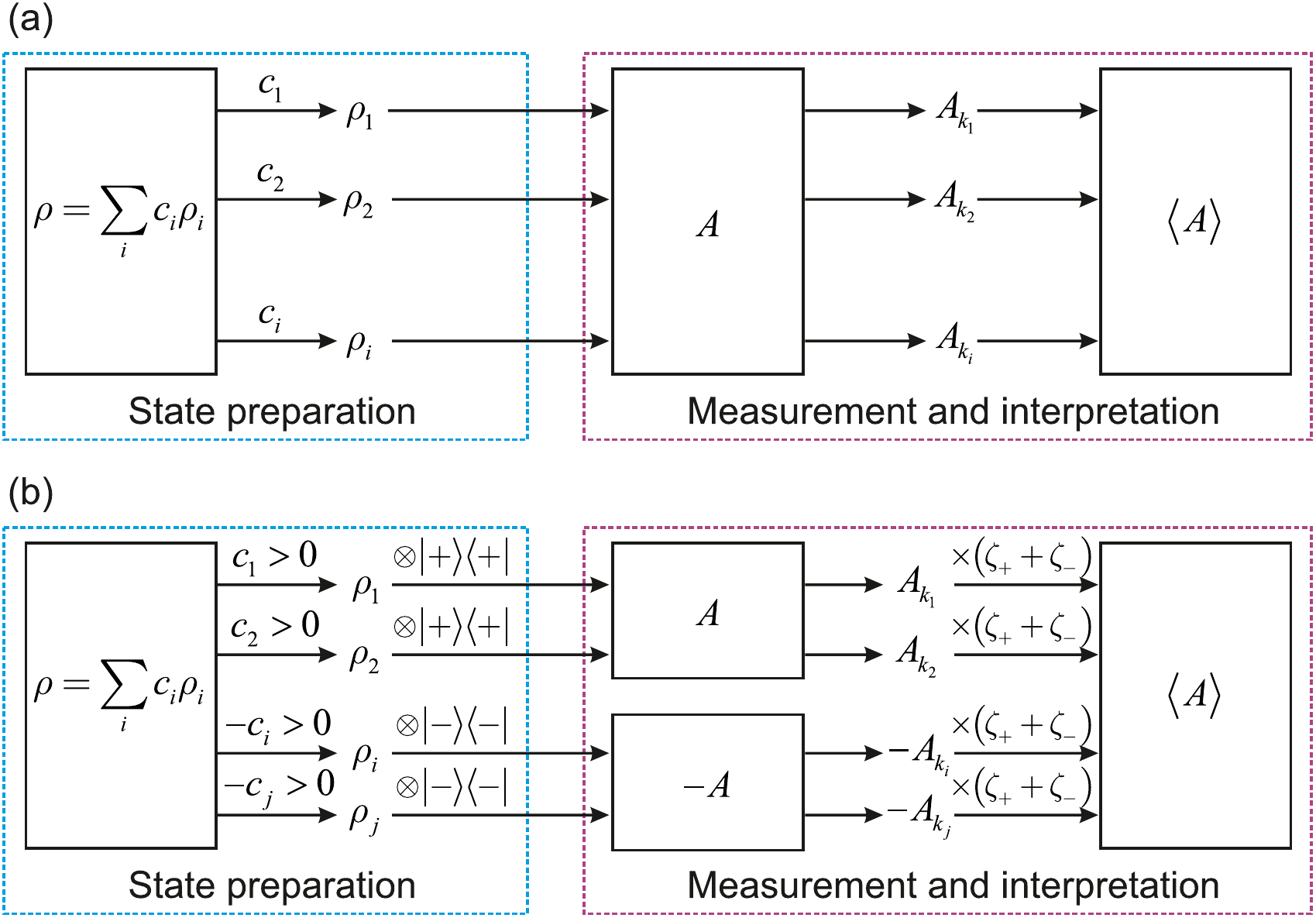}
	\caption{The measurement of (a)~a regular quantum statistical mixture and (b)~a general emulation representation stated in~(\ref{eqn:representation}). In both kinds of state preparation: the component $\rho_i$ is produced by the source with probability proportional to $|c_i|$. In the case of a statistical mixture state, estimation of the expectation value $\left<A\right>$ of an observable $A$ requires no knowledge about how $\rho$ is prepared $(\rho_i)$. In the \emph{emulation} of $\rho$ that requires some negative $c_i$'s, classical information concerning state preparation is necessary: The observer measures either $A$ or $-A$ depending on the ancillary measurement of $\rho_\textsc{c}$ with $B$.}
	\label{fig:schemes}
\end{figure}

To demonstrate feasibility of our approach, we show how to emulate  fundamental quantum results such as antibunching, violation of Bell inequalities, and witnessing non-classicality, and provide estimation of classical resources required for it. Our scheme has important practical applications. For example, it allows one to test quantum effects  when it is problematic or too expensive to generate ``true'' non-classical states, such as the NOON states with several photons, or, for example, for low wavelengths (such as the microwave spectral region, where one needs implementing superconducting circuits \cite{radars}).  

The outline of the paper is as follows. In the Section II we discuss the ideology of our emulation scheme and show how one includes classical information on the state preparation in the emulation set-up. In the third Section we discuss high-fidelity representations of several few-photon states including Fock and NOON-states. In the Section IV we describe the way of witnessing non-classicality of emulated states with realistic single-photon detectors. In ther Sections V, VI and VIII we discuss demonstration of the Hong-Ou-Mandel effect,  phase estimation and Bell testing with our emulated states. 

\section{Emulation scheme}

Now let us demonstrate how it is possible to emulate an arbitrary measurement over an arbitrary quantum state  using info about the state preparation. We assume emulating measurement of the observable $A$. Let us consider a state described by the density matrix $\rho_{\mathrm{true}}$ and  approximate it by the operator $\rho$ represented through a set of probe states $\{\rho_i\}$ in the following way:
\begin{equation}
\rho = \sum_j c_j \rho_j, \quad \sum_j c_j=1,
\label{eqn:representation}
\end{equation}
where the coefficients $c_j$ can be negative for non-orthogonal $\rho_j$. The emulation of the measurement results is faithful if for an arbitrary small $\epsilon>0$ we can find such representation (\ref{eqn:representation}) that 
\[
|\mathrm{Tr}\{A(\rho-\rho_{\mathrm{true}})\}|<\epsilon.\] 
For any $A$ with limited eigenvalues, $|\lambda_i| \le M$ ($M = 1$ if a photon(s) detection probability is measured as in the examples below), the faithfulness condition is satisfied when the representation fidelity is high enough (see Sec. 1 of the Appendix): 
\[
F(\rho, \rho_\text{true}) = [ \operatorname{Tr} \sqrt{\sqrt{\rho}\rho_\text{true}\sqrt{\rho}}]^2 \ge 1 - \epsilon^2 / (4 M^2)
\]
It is well-known that one can always build such a representation with a mixture of coherent states projectors, either by {a} continuous Glauber-Sudarshan P representation  \cite{PhysRevLett.10.277,PhysRev.131.2766}, or its ``coarse-grained'' discrete version \cite{PhysRevLett.16.534,Lobino563}.  Recently in papers discussing the ``data pattern'' approach it was shown how to  approximate a given state (or its projection on some subspace) with high fidelity by a finite (and rather small) number of coherent state projectors on some predefined lattice \cite{PhysRevLett.105.010402,Mogilevtsev_2013,Motka_2017,reut2017}.  

To incorporate classical {information on the signs} of $c_j$,
we build the following combined state of our signal and the two-state ancilla labeling the prepared states:
\begin{equation}
\label{combined}
\rho_{c} = \frac{\zeta_+}{\zeta_++\zeta_-}\rho_+\otimes |+\rangle\langle+|+\frac{\zeta_-}{\zeta_++\zeta_-}\rho_-\otimes |-\rangle\langle-|,
\end{equation}
where 
\[
\zeta_{+(-)}=\sum\limits_{c_j>0(<0)}|c_j|, \quad \rho_{+(-)}=\zeta_{+(-)}^{-1}\sum\limits_{c_j>0(<0)}|c_j|\rho_j, 
\]
and the two mutually orthonormal ancilla states $|\pm\rangle$ encode classical information about the sign of the coefficient $c_j$ before the sampled signal state $\rho_j$. Notice that all the weights in the mixture of combined probe states (\ref{combined}) are positive. To utilize the knowledge about the state preparation for measuring the observable $A$, we suggest measuring the combined observable  
$A\otimes B$, where the ancilla observable allowing to infer the info about the state preparation is  
\begin{equation}
B=(\zeta_++\zeta_-)(|+\rangle\langle+|-|-\rangle\langle-|). 
\label{b}
\end{equation}
Thus, up to the accuracy of the representation,
\[\langle A \rangle=\mathrm{Tr}\{(A\otimes B)\rho_c\}.\] 
Notice that {for the positive-weighted mixtures} of coherent-state projectors, the state (\ref{combined}) becomes trivial, and the measurement procedure is the same as for a usual, preparation-indifferent  measurement. The scheme for measuring $\rho_\textsc{c}$ defined in Eq.~(\ref{combined}) for {general mixed states} and emulated states are shown in Fig.~\ref{fig:schemes}. As follows from Eq.(\ref{combined}),  the described measurement can be realized in the following simple way. One samples the probes $\rho_j$ according to the probability distribution 
\[p_j = {|c_j|}/(\zeta_+ + \zeta_-),\] 
labels each probe state by the ancilla state $|+\rangle$ or $|-\rangle$ depending on $\operatorname{sgn}(c_j)$, and performs the measurement of the observable $A$ on the signal state and $B$ on the ancilla. If the $k$-th sample is the probe $\rho_{j_k}$, let us denote the particular measurement result of the observable $A$ as $A_k$ and the classical weight (measurement result for $B$) as $B_k$. For $N$ samples, we calculate the following combination of the measurement results:
\begin{equation}
        \label{eqn:measurement_result}
        \langle A \rangle_N =\frac{1}{N} \sum_k A_k B_k = \frac{\zeta_+ + \zeta_-}{N} \sum_k \operatorname{sgn}(c_{j_k}) A_k,
    \end{equation}
with 
$\left<A\right>=\langle A \rangle_N +O(1/\sqrt{N})$ (for details, please, see Sec. 2 of the Appendix). Eqs. (\ref{eqn:representation})--(\ref{eqn:measurement_result}) show that using just coherent states, it is possible to emulate results of any measurements on the quantum state, $\rho_{\mathrm{true}}$, with arbitrary precision. However, one needs to pay for it by the necessity of additional  measurements of the ancilla resulting in positive and negative weights $B_k$, and that leads  to increase in statistical errors. Indeed, from Eqs. (\ref{eqn:representation},\ref{combined},\ref{b}) and using $\zeta_+- \zeta_-=1$, one can get the following expressions for the variances, 
\begin{eqnarray}
\nonumber
\Delta_{AB}=(\zeta_++\zeta_-)(\zeta_+\mathrm{Tr}\{\rho_+A^2 \} + \zeta_-\mathrm{Tr}\{\rho_-A^2 \})-\left<A\right>^2,\\
\nonumber
\Delta_{A}=(\zeta_+-\zeta_-)(\zeta_+\mathrm{Tr}\{\rho_+A^2 \} - \zeta_-\mathrm{Tr}\{\rho_-A^2 \})-\left<A\right>^2,
\end{eqnarray}
where the variance $\Delta_{AB}$ is of $A\otimes B$ evaluated with the mixed state $\rho_{c}$, and the variance $\Delta_A$ is of the observable $A$ evaluated with $\rho$. Thus, one has for the difference 
\begin{equation}
    \label{eqn:noise}
    \begin{gathered}
    \Delta_{AB}- \Delta_A = 2 \zeta_+ \zeta_- \left( \mathrm{Tr}\{\rho_+A^2 \} + \mathrm{Tr}\{\rho_-A^2 \}\right) \ge 0.
    \end{gathered}
\end{equation}
The price for the ability to model quantum states by {mixtures of non-negative-P-function states} is a larger number of the measurement runs for getting the same statistical error ( some consideration on the sampling accuracy in non-classical quantum-state emulation are provided in Sec. 2 of the Appendix). However, as we shall see below, the emulation procedure might be quite economical in terms of used  resources. 

\section{Emulation feasibility}

It is already established that for few-photon and few-mode non-classical states one can achieve higher than $0.99$ fidelity of the approximation (\ref{eqn:representation}) and reproduction of the experiment results with  just few tens of the probe coherent states \cite{PhysRevLett.105.010402,Mogilevtsev_2013,reut2017}. For few-photon Fock states it is possible to develop quite economical representations in terms of phase-averaged coherent states. Let us show here how it is possible to represent even entangled states using few phase-averaged coherent states and a set of  simple optical devices such as beam-splitters and phase-shifters. 

\begin{figure}[tp]
\includegraphics[width=\linewidth]{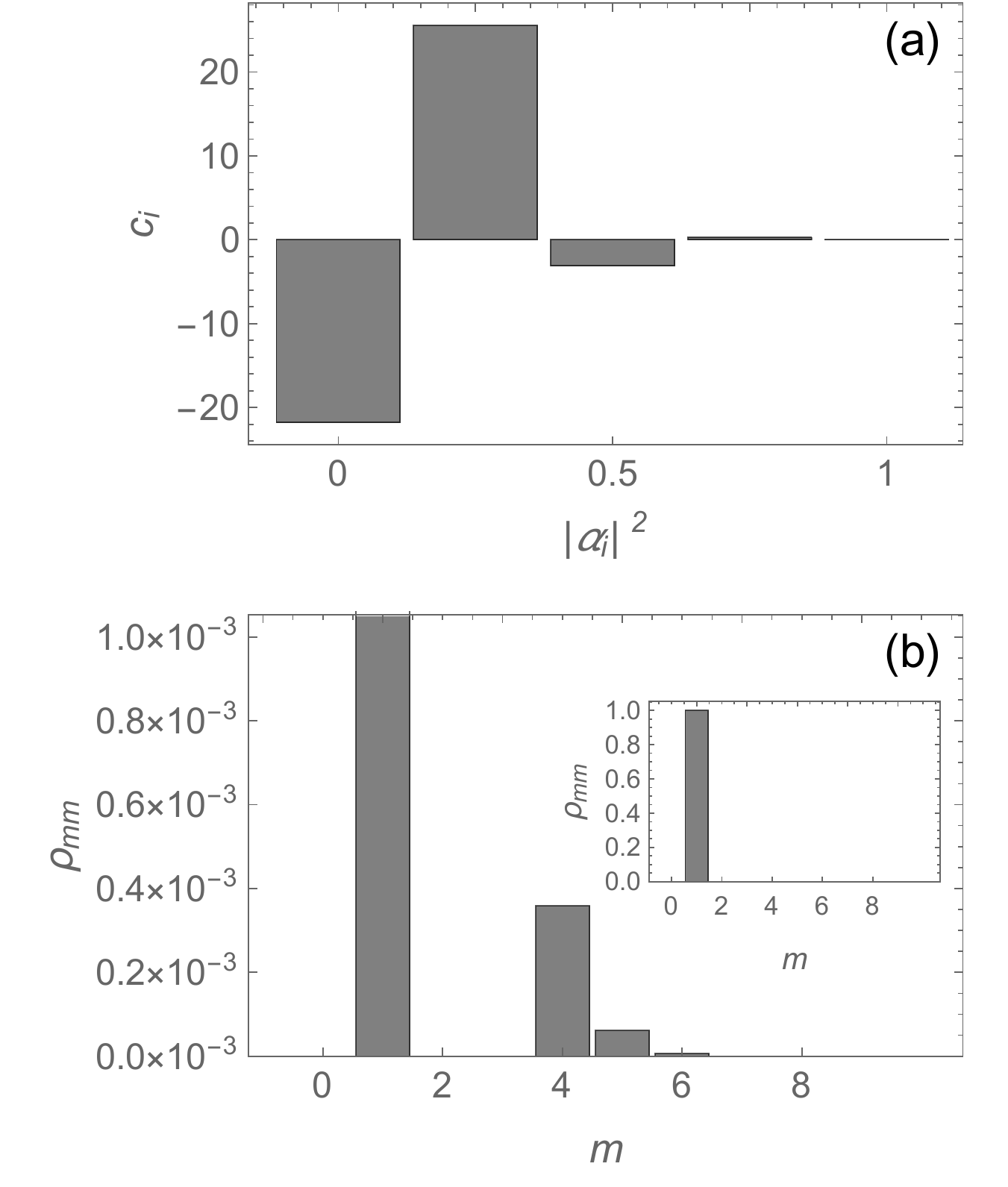}
\caption{Representation of the single-photon state in terms of phase-averaged coherent states: decomposition coefficients (a), diagonal elements of the optimal linear combination of the coherent states, shown in different scales in the main plot and the inset (b).}
\label{fig:single-photon}
\end{figure}

\begin{figure}[tp]
\includegraphics[width=\linewidth]{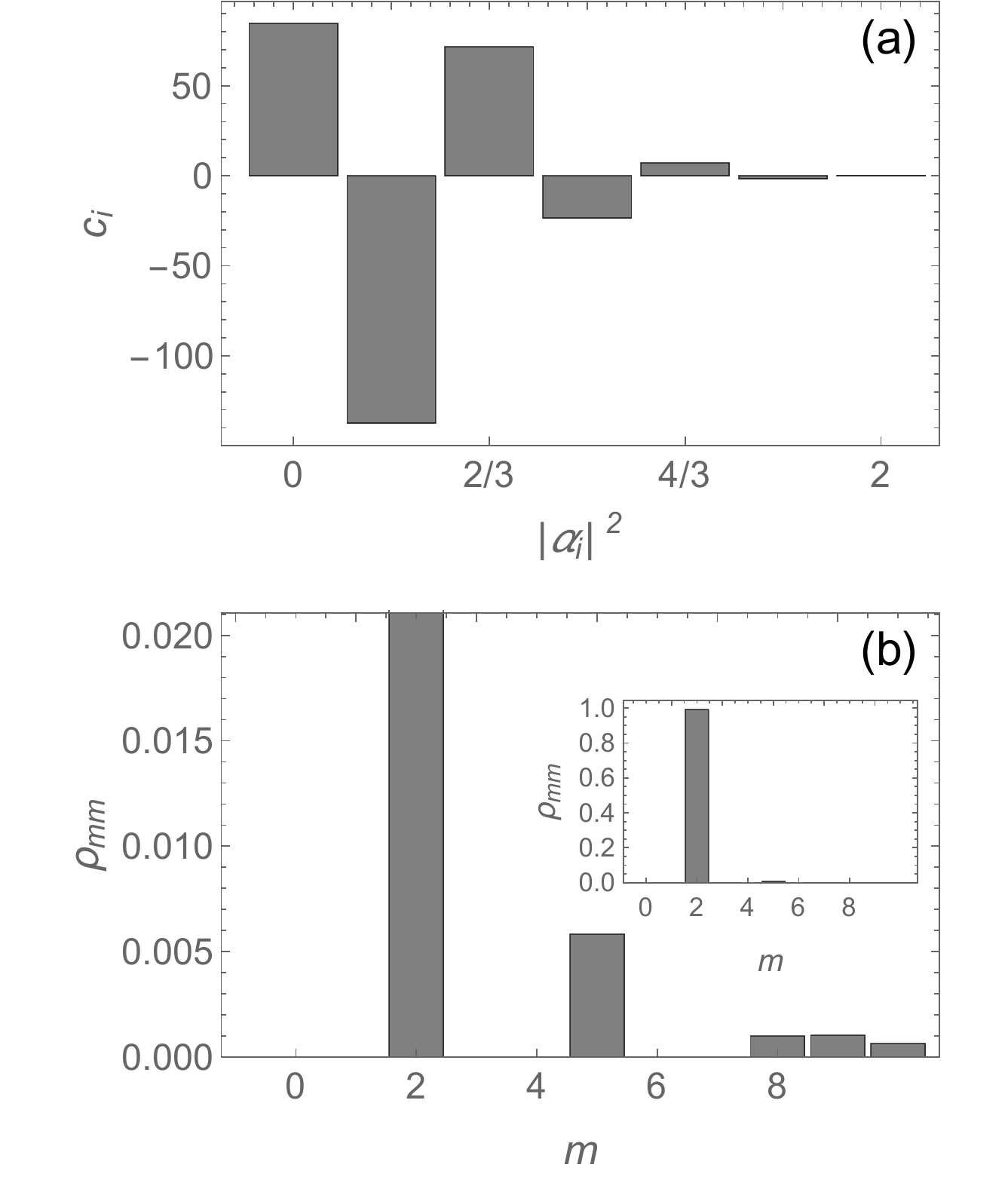}
\caption{Representation of the two-photon state in terms of phase-averaged coherent states: decomposition coefficients (a) and diagonal elements of the optimal linear combination of the coherent states, shown in different scales in the main plot and the inset (b).}
\label{fig:two-photon}
\end{figure}

\subsection{Single-photon and two-photon states emulation}

First of all, let us show examples of  representations for single-photon and two-photon states. 
In general, given a fixed set of $\rho_j$'s, the coefficients $c_j$, which approximate $\rho_\text{true}$ in the best way, can be determined by solving the following numerical problem:
\begin{align}
&\,\max_{\{c_j\}}\,F(\rho_\text{true},\rho)\nonumber\\
&\,\text{subject to: }\sum_jc_j=1,\,\,\rho\geq0\,,
\label{eqn:sdp}
\end{align}
where the fidelity 
\[F(\rho_\text{true},\rho) = (\mathrm{Tr}\left\{\sqrt{\sqrt{\rho_\text{true}}\,\rho\sqrt{\rho_\text{true}}}\right\})^2\] 
is maximized over the coefficients $c_j$ conditioned on the positive semidefiniteness of $\rho$~\cite{reut2017}. The solution to this problem can be found using semidefinite programming.

When $\rho_\text{true}=\ket{1}\bra{1}$ is the single-photon state, the {set of probes} can be chosen in the form of 5 phase-averaged coherent states with amplitudes $\alpha_i = 0$, 0.25, 0.5, 0.75, 1 (Fig.~\ref{fig:single-photon}a):
\begin{multline}
\label{eqn:probe}
\rho_j = \overline{|\alpha_j \rangle \langle \alpha_j |} \equiv \frac{1}{2 \pi} \int\limits_0^{2\pi} d\varphi |\alpha_j e^{i \varphi}\rangle \langle \alpha_i e^{i \varphi}| \\{} = \sum_{n=0}^\infty \frac{|\alpha_j|^{2n}}{n!}e^{-|\alpha_j|^2}|n\rangle \langle n|.
\end{multline}
The semidefinite program in \eqref{eqn:sdp} produces the resulting optimal coefficients $c_j=-21.8,25.6,-3.1,0.33,-0.0028$. The fidelity of the constructed representation for the single photon state exceeds 0.9996 (Fig.~\ref{fig:single-photon}b).

 Similarly, as it is shown in Fig.~\ref{fig:two-photon}, just seven phase-averaged coherent states allow to represent two-photon Fock states with the representation fidelity exceeding 0.998. 

\subsection{NOON state emulation}

Now let us demonstrate a potential of our approach by emulating entangled bipartite states, namely, the NOON states, 
\begin{equation}
\label{noon}
|\Psi_N\rangle_{ab} = \frac{1}{\sqrt 2} \left(|N\rangle_a |0\rangle_b - |0\rangle_a|N\rangle_b\right),
\end{equation}  
with only the phase-averaged coherent states, and with practically the same effort as the $(N-1)$-photon Fock states (for $N \ge 2$).  For $N=1,2$ it is trivially accomplished  by 50/50 beam-splitting and single-photon/vacuum or two single-photon inputs. Let us show that beam-splitting and phase-shifting allows easy producing the NOON states with an arbitrary N. 

We have already shown above that Fock states can be decomposed in terms of phase-averaged coherent states and require affordable resources. Linear optical transformation of the Fock states correspond to trivial arithmetic operations with the amplitudes of the coherent states, used for their representation. Therefore, finding a way to represent a NOON state as a result of some linear operations applied to Fock states will be sufficient for construction of its efficient decomposition.

First, let us consider the linear optical transformation
\begin{equation}
\label{eqn:beamsplitter}
a^\dag \rightarrow (a^\dag + e^{i\theta} b^\dag)/\sqrt 2,\quad b^\dag \rightarrow (a^\dag - e^{i\theta} b^\dag)/\sqrt 2
\end{equation}
applied to the 2-mode Fock state $|n\rangle_a |m\rangle_b$ with $n + m = N$. The density matrix of the resulting state is
\begin{multline}
\label{eqn:beamsplitter_rho}
\rho(n,m,\theta) = \sum_{j = 0}^N \sum_{l = 0}^N R_{nm}^{(j)} R_{nm}^{(l)} e^{i \theta (l - j)} \\{} \times |j\rangle_a \langle l| \otimes |N - j \rangle_b \langle N - l |,
\end{multline}
where
\begin{equation}
\label{eqn:A_nm_j}
R_{nm}^{(j)} = \sum_{k = \max(0, j - m)}^{\min(n, j)} \frac{\sqrt{n! m! j! (N - j)!} (-1)^{m - j + k}}{k! (n-k)! (j - k)! (m - j + k)!} 2^{-N/2}.
\end{equation}
In the symmetric case $n = m = N / 2$, only even indices $j = 2 s$ yield non-zero coefficients (it can be considered as a generalization of Hong-Ou-Mandel effect):
\begin{equation}
\label{eqn:A_nm_j_sym}
R_{nn}^{(2s)} = \frac{\sqrt{(2s)! (N - 2s)!} (-1)^{n - s}}{s! (n-s)!} 2^{-N/2}.
\end{equation}

To generate the target NOON state, we need to remove all the terms from Eq.~(\ref{eqn:beamsplitter_rho}), except for those with $j$ and $l$ equal to 0 or $N$. Here, we can use the equality
\begin{equation}
\label{eqn:sum_exp}
\frac{1}{N}\sum_{k = 0}^{N-1} e^{i \frac{2 \pi jk}{N}} = \left[\begin{array}{cc}
     1, & k = 0, \pm N, \pm 2 N, \ldots \\
     0,& \text{otherwise}. 
\end{array} \right.
\end{equation}

Therefore,
\begin{multline}
\label{eqn:beamsplitter_rho_sum}
\frac{1}{N}\sum_{k=0}^{N-1} \rho(n,m,\theta_0 + 2 \pi k /N) \\{}= \frac{N!}{n! m!}2^{-N+1} |\Psi_N(\theta_0, m)\rangle_{ab} \langle \Psi_N(\theta_0, m)| \\{} + \sum_{j = 1}^{N - 1} \left(R_{nm}^{(j)}\right)^2 |j\rangle_a \langle j| \otimes |N - j \rangle_b \langle N - j|,
\end{multline}
where 
\begin{equation}
\label{eqn:Psi_theta}
|\Psi_N(\theta_0,m)\rangle_{ab} = \frac{1}{\sqrt 2}\left(|N\rangle_a|0\rangle_b + (-1)^m e^{i N\theta_0} |0 \rangle_a |N \rangle_b\right).
\end{equation}
By the choice $\theta_0 = \pi / N$ for even $m$ and $\theta_0 = 0$ for odd $m$, one can ensure that the first term of Eq.~(\ref{eqn:beamsplitter_rho_sum}) corresponds to the target state (\ref{noon}): $|\Psi_N(\theta_0,m)\rangle_{ab} = |\Psi_N\rangle_{ab}$.

Finally, the target state can be expressed from Eq.~(\ref{eqn:beamsplitter_rho_sum}):
\begin{multline}
\label{eqn:decomposition_N}
|\Psi_N\rangle_{ab} \langle \Psi_N | = \frac{n!m!2^{N-1}}{N!} \Biggl[\frac{1}{N}\sum_{k=0}^{N-1} \rho\left(n,m,\theta_0 + \frac{2 \pi k}{N}\right) \\ {} - \sum_{j = 1}^{N - 1} \left(R_{nm}^{(j)}\right)^2 |j\rangle_a \langle j| \otimes |N - j \rangle_b \langle N - j| \Biggr].
\end{multline}
As discussed above, the derived expression implies that emulation of the NOON-state is not much more complex than emulation of the Fock state with $N - 1$ photons.

When $N$ is even, the procedure can be simplified. Eq.~(\ref{eqn:A_nm_j_sym}) implies that only even multipliers $(l - j)$ of the parameter $\theta$ are present in Eq.~(\ref{eqn:beamsplitter_rho}) if $n = m = N/2$. Therefore, $\rho(n,n,\theta + \pi) = \rho(n,n,\theta)$, and the summation over $j$ can be limited by $N/2 - 1$ instead of $N - 1$:
\begin{multline}
\label{eqn:decomposition_N_even}
|\Psi_N\rangle_{ab} \langle \Psi_N | \\ = \frac{(n!)^2 2^{N-1}}{N!} \Biggl[\frac{2}{N}\sum_{k=0}^{N/2-1} \rho\left(n,m,\theta_0 + \frac{2 \pi k}{N}\right) \\ {} - \sum_{k = 1}^{N/2 - 1} \left(R_{nm}^{(2k)}\right)^2 |2k\rangle_a \langle 2k| \otimes |N - 2k \rangle_b \langle N - 2k| \Biggr].
\end{multline}
In comparison with Eq.~(\ref{eqn:decomposition_N}), the derived expression contains almost twice smaller number of terms and requires emulation of Fock states with up to $\max(N / 2, N - 2)$ photons only.

The expressions (\ref{eqn:decomposition_N}) and (\ref{eqn:decomposition_N_even}) have exactly the same form as required by Eq.~(\ref{eqn:representation}). Therefore, the emulation of the target NOON-state can be performed as discussed above, but applied in two steps. First, one randomly chooses one of the states from the right-hand side of Eq.~(\ref{eqn:decomposition_N}) or (\ref{eqn:decomposition_N_even}) with their probabilities being proportional to the decomposition coefficients. Then, the selected state is emulated {"classically"} according to its decomposition in terms of coherent states. Let us consider the emulation of the states from the right-hand side of Eqs.~(\ref{eqn:decomposition_N}) and (\ref{eqn:decomposition_N_even}) in more details.

Suppose that the Fock states $|n\rangle$ can be approximated by linear combinations of phase-averaged coherent states in the following way:
\begin{equation}
\label{eqn:representation_Fock}
|n\rangle \langle n| \approx \sum_i c_{ni} \overline{|\alpha_{ni} \rangle \langle \alpha_{ni} |}.
\end{equation}
Therefore, 2-mode Fock states can be emulated as
\begin{equation}
\label{eqn:representation_double_Fock}
|n\rangle_a \langle n| \otimes |m\rangle_b \langle m| \approx \sum_{i,j} c_{ni} c_{mj} \overline{|\alpha_{ni} \rangle_a \langle \alpha_{ni} |} \otimes \overline{|\alpha_{mj} \rangle_b \langle \alpha_{mj} |}.
\end{equation}

Finally, the states $\rho(n,m,\theta)$ can be decomposed as
\begin{equation}
\label{eqn:representation_beamsplitter}
\rho(n,m,\theta) = \sum_{i,j} c_{ni} c_{mj} \rho_{ij}(n,m,\theta),
\end{equation}
where
\begin{multline}
\label{eqn:representation_beamsplitter_coherent}
\rho_{ij}(n,m,\theta) = \frac{1}{2\pi} \int_0^{2\pi} d \varphi_1 \int_0^{2\pi} d \varphi_2  \\ {} \times   \left|\frac{|\alpha_{ni}| e^{i\varphi_1} + |\alpha_{mj}| e^{i \varphi_2}}{\sqrt 2}\right\rangle_a \left\langle\frac{|\alpha_{ni}| e^{i\varphi_1} + |\alpha_{mj}| e^{i \varphi_2}}{\sqrt 2}\right| \otimes {} \\ \left|e^{i\theta}\frac{|\alpha_{ni}| e^{i\varphi_1} - |\alpha_{mj}| e^{i \varphi_2}}{\sqrt 2}\right\rangle_b \left\langle e^{i\theta}\frac{|\alpha_{ni}| e^{i\varphi_1} - |\alpha_{mj}| e^{i \varphi_2}}{\sqrt 2}\right|.
\end{multline}
Technically, generation of a phase-averaged coherent state corresponds to generation of a coherent state with the given amplitude and addition of a random uniformly distributed phase shift. I.e. to emulate the state $\rho(n,m,\theta)$, one chooses the pair of indices $(i,j)$ with the probabilities proportional to $c_{ni} c_{mj}$, then choose two random phases $\varphi_1, \varphi_2 \in [0, 2 \pi)$, and finally generates the 2-mode coherent state according to the integrand of Eq.~(\ref{eqn:representation_beamsplitter_coherent}).

For $N = 1$, $n = 1$, and $m = 0$, Eq.~(\ref{eqn:decomposition_N}) yields
\begin{equation}
\label{eqn:NOON_1}
|\Psi_-\rangle_{ab} \langle \Psi_- | = \rho(1,0,\pi), 
\end{equation}
which completely agrees with the previously obtained results.

For $N = 2$, $n = 1$, and $m = 1$, one can use Eq.~(\ref{eqn:decomposition_N_even}) to obtain the representation
\begin{equation}
\label{eqn:NOON_2}
|\Psi_2\rangle_{ab} \langle \Psi_2 | = \rho(1,1,0), 
\end{equation}
known from Hong-Ou-Mandel effect.

For $N = 3$, $n = 2$, $ m = 1$ and $N = 4$, $n = 2$, $m = 2$ the results are
\begin{multline}
\label{eqn:NOON_3}
|\Psi_3\rangle_{ab} \langle \Psi_3 | = \frac{4}{9} \left[\rho(2,1,0) + \rho(2,1, 2\pi / 3) + \rho(2,1,4 \pi / 3) \right] \\{} - \frac{1}{6}\left(|1\rangle_a\langle 1| \otimes |2\rangle_b\langle 2| +|2\rangle_a\langle 2| \otimes |1\rangle_b\langle 1| \right)
\end{multline}
and
\begin{multline}
\label{eqn:NOON_4}
|\Psi_4\rangle_{ab} \langle \Psi_4 | = \frac{2}{3} \left[\rho(2,2,\pi / 4) + \rho(2,2, 3\pi / 4) \right] \\{} - \frac{1}{3} |2\rangle_a\langle 2| \otimes |2\rangle_b\langle 2|
\end{multline}
respectively.

So, we can see that for emulation of NOON-states with $N\leq 4$ it is sufficient to implement representation of just a single and two-photon states considered in the previous Subsection. The final fidelity of NOON-states decomposition with the considered representations are listed in Table~\ref{tab:NOON}.

\begin{table}
    \centering
    \begin{tabular}{|c|c|c|c|c|}
    \hline
         & $N = 1$ & $N = 2$ & $N = 3$ & $N = 4$ \\ \hline
     Fidelity    &  0.9996 & 0.9992 & 0.99 & 0.982 \\ \hline 
     $\zeta_+ + \zeta_-$    &  51 & $2.6\times 10^3$  & $2.8\times 10^4$ & $1.8 \times 10^5$ \\ \hline 
    \end{tabular}
    \caption{Results for "classical" emulation of NOON-states}
    \label{tab:NOON}
\end{table}

As mentioned above, the derived expression implies that emulation of the NOON-state is not much more complex than emulation of the Fock state with $N - 1$ photons. 

Below it is shown how to perform phase estimation with emulated NOON states. By experimenting with such NOON states, one can confirm the expected quantum effects without the troubles related to the generation and preservation of complex non-classical quantum states. Here it is also useful to mention that decoherence very quickly deteriorates metrological advantage expected from the true NOON states \cite{PhysRevLett.79.3865}.

One should emphasize that the measurement-oriented representation might be  more economical that the one discussed above. Indeed, {for faithful emulation of measuring} the observable $A$ diagonal in the Fock-state basis, it is sufficient to emulate Fock-state mixtures instead of superpositions. Also, the complexity of representation is not connected with the state energy, but rather with the number of required basis states. For example, to represent the ``cat-state''  $|\alpha\rangle+|\alpha+\delta\rangle$ with $|\delta|\ll|\alpha|$ and an arbitrary $\alpha$, one might need only a few coherent projectors with amplitudes close to $\alpha$. 

Also, the choice of {state mixtures} for emulation is not limited to coherent states or their phase-averaged version. One might guess that the representations akin to ones discussed above can be developed with other states, and tailored for a particular measurement. For example, thermal states were used to represent Fock states in Ref.\cite{Mogilevtsev_2013}. Some considerations on choosing the best {mixtures} for a particular measurements are given in Sec. 4 of the Appendix. 

\begin{figure}[tp]
\includegraphics[width=\linewidth]{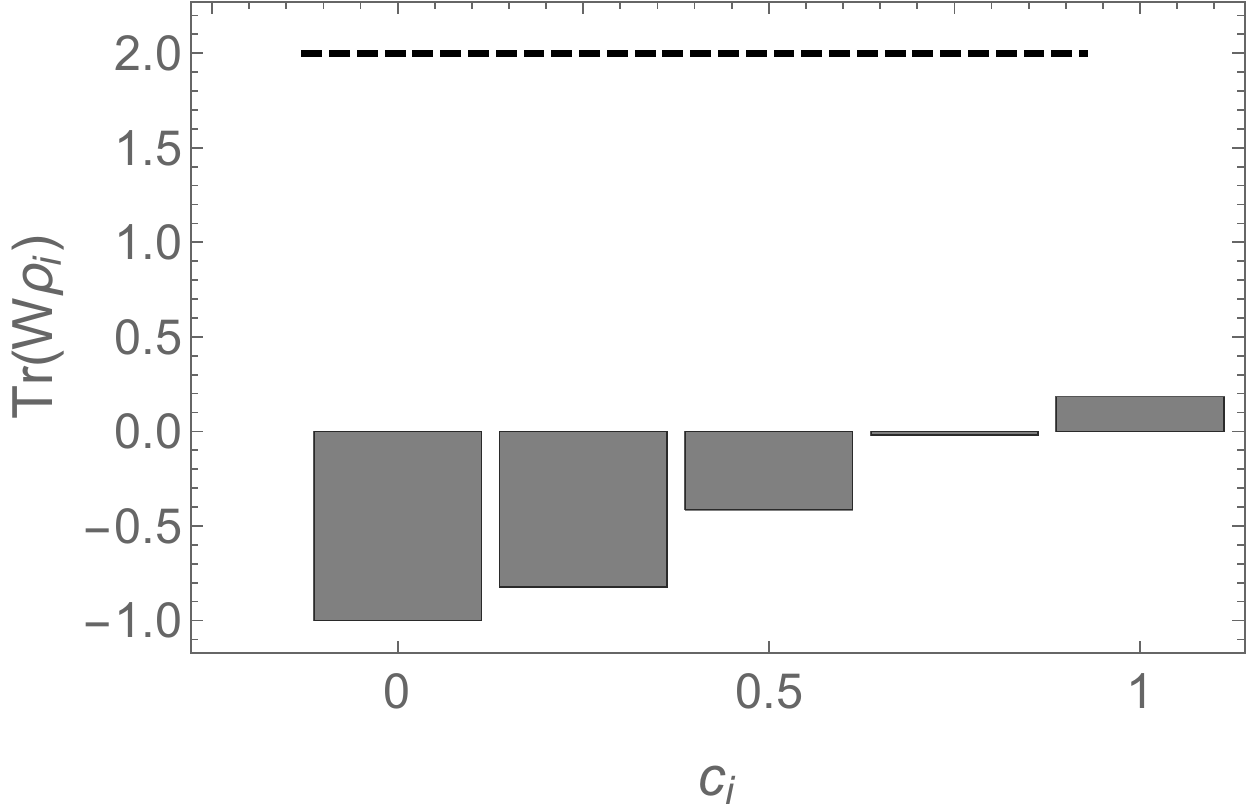}
\caption{The values of non-classicality witness operator $W$ (Eq.~(\ref{eqn:witness_single_photon})) for the coherent states of the single-photon state representation depicted in Fig.~\ref{fig:single-photon}a. The dashed horizontal line shows the expectation value of the non-classicality witness operator for the single-photon state. }
\label{fig:wit}
\end{figure}

\section{Non-classicality witness}

Now let us demonstrate how can one certify non-classicality of the emulation results.

To prove non-classicality of a given state (or a class of states), one can construct a witness operator $W$, find the classical limit 
\begin{equation}
    \label{eqn:witness_limit}
    W_0 = \max_{\rho' {\in\mathcal{C}}} \mathrm{Tr}\{W\rho'\},
\end{equation}
{where $\mathcal{C}$ is the set of non-negative-P-function states,} and check that the investigated state $\rho$ yields
\begin{equation}
    \label{eqn:witnessing_nonclassicality}
    \Tr{W \rho} > W_0.
\end{equation}

For example, to prove non-classicality of the single-photon state, one can build the following witness operator:
\begin{equation}
    \label{eqn:witness_single_photon}
    W = 2 |1\rangle \langle 1| - |0 \rangle \langle 0| - |2 \rangle \langle 2|.
\end{equation}

Using the {the developed representation for} $\rho'$ and taking into account diagonality of the operator $W$ in Fock state basis, one can show that
\begin{equation}
    \label{eqn:maximization}
    \begin{gathered}
    W_0 = \max_{\rho' {\in\mathcal{C}}} \Tr{W\rho'} \\
    {} = \max_{P: P(\alpha) \ge 0} \int d^2 \alpha P (\alpha) W(|\alpha|),
    \end{gathered}
\end{equation}
where 
\begin{equation}
    \label{witness_function}
    W(|\alpha|) = \langle \alpha | W | \alpha \rangle = \left(2|\alpha|^2 - 1 -\frac{|\alpha|^4}{2} \right) e^{-|\alpha|^2},
\end{equation}
and the normalization condition holds:
\begin{equation}
    \label{eqn:normalization}
    \int d^2\alpha P(\alpha) = 1.
\end{equation}

{A non-negativity of $P(\alpha)$} implies that the maximal classical value $W_0$ corresponds to the maximum of the function $W(|\alpha|)$, which equals
\begin{equation}
    \label{eqn:witness_0}
    W_0 = \max_\alpha W(|\alpha|) = 0.206
\end{equation}
and is reached for the coherent state with the amplitude $|\alpha^{(0)}| = 1.134$.

For the single-photon state, the expectation value of the witness operator equals $\langle 1 | W | 1 \rangle = 2 > W_0$.

The mean value $\langle W \rangle$ for the emulated  state (\ref{eqn:probe}) equals 1.9992, which clearly exceeds the classical limit $W_0 = 0.206$. On the other hand, Fig.~\ref{fig:wit} shows that the values $\Tr{W \rho_j}$ fit into the classical region $[-1, W_0]$ for all $j$. The two reasons for the final result exceeding the classical limit are:
\begin{itemize}
    \item minus sign for certain $\rho_j$: classical maximum of $-W$ is $1 > 0.206$ (but still less than 2);
    \item the measurement results are multiplied by the factor $\zeta_+ + \zeta_- = 50.8$.
\end{itemize}

The excess variance of a single-trial measurement (given by Eq. (5) of the main text) is $2.0 \times 10^3$. Therefore, to demonstrate the non-classicality reliably, one needs of about $10^5$ copies of the state.

\begin{figure}[tp]
\includegraphics[width=\linewidth]{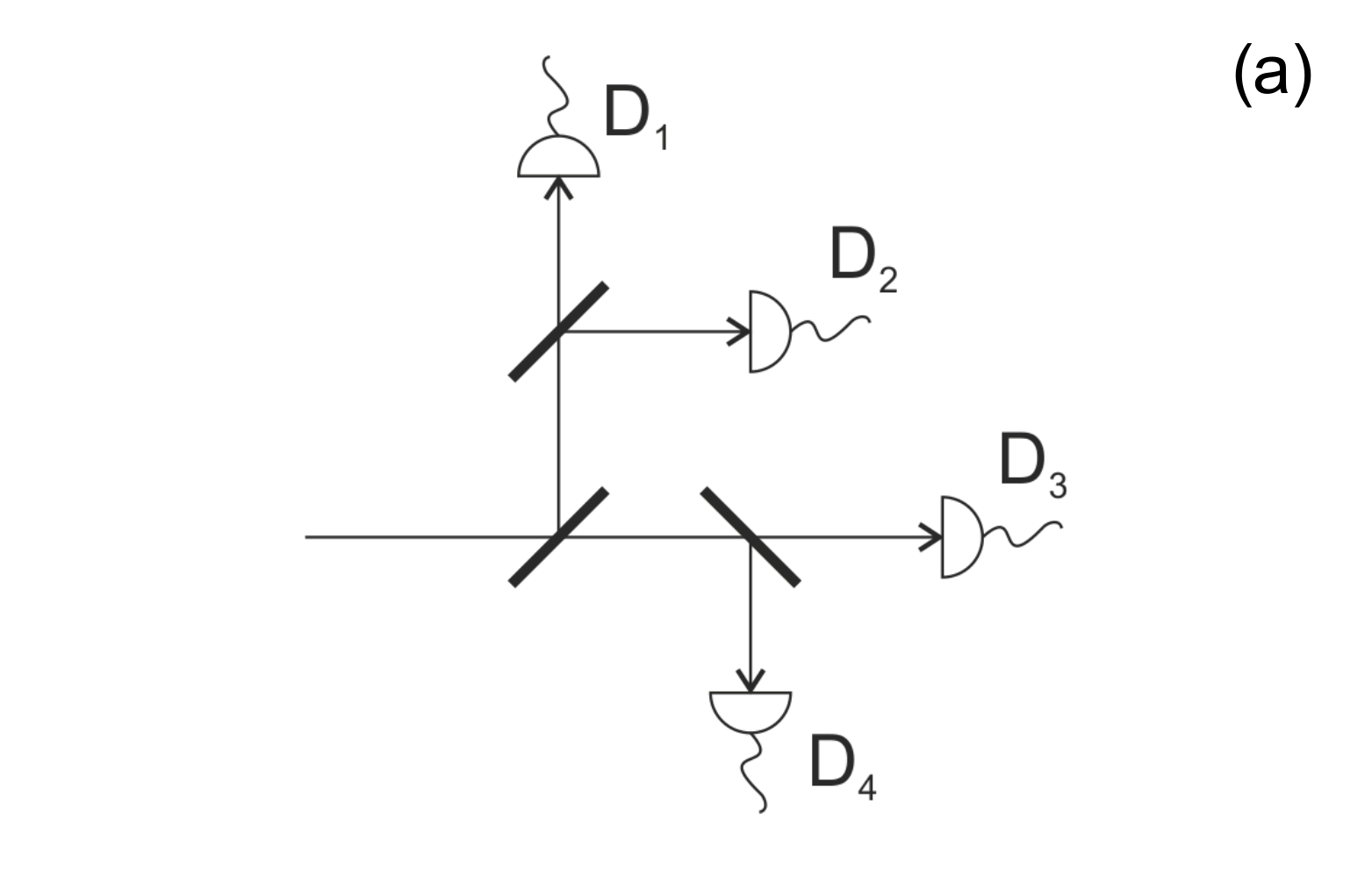}\\
\includegraphics[width=\linewidth]{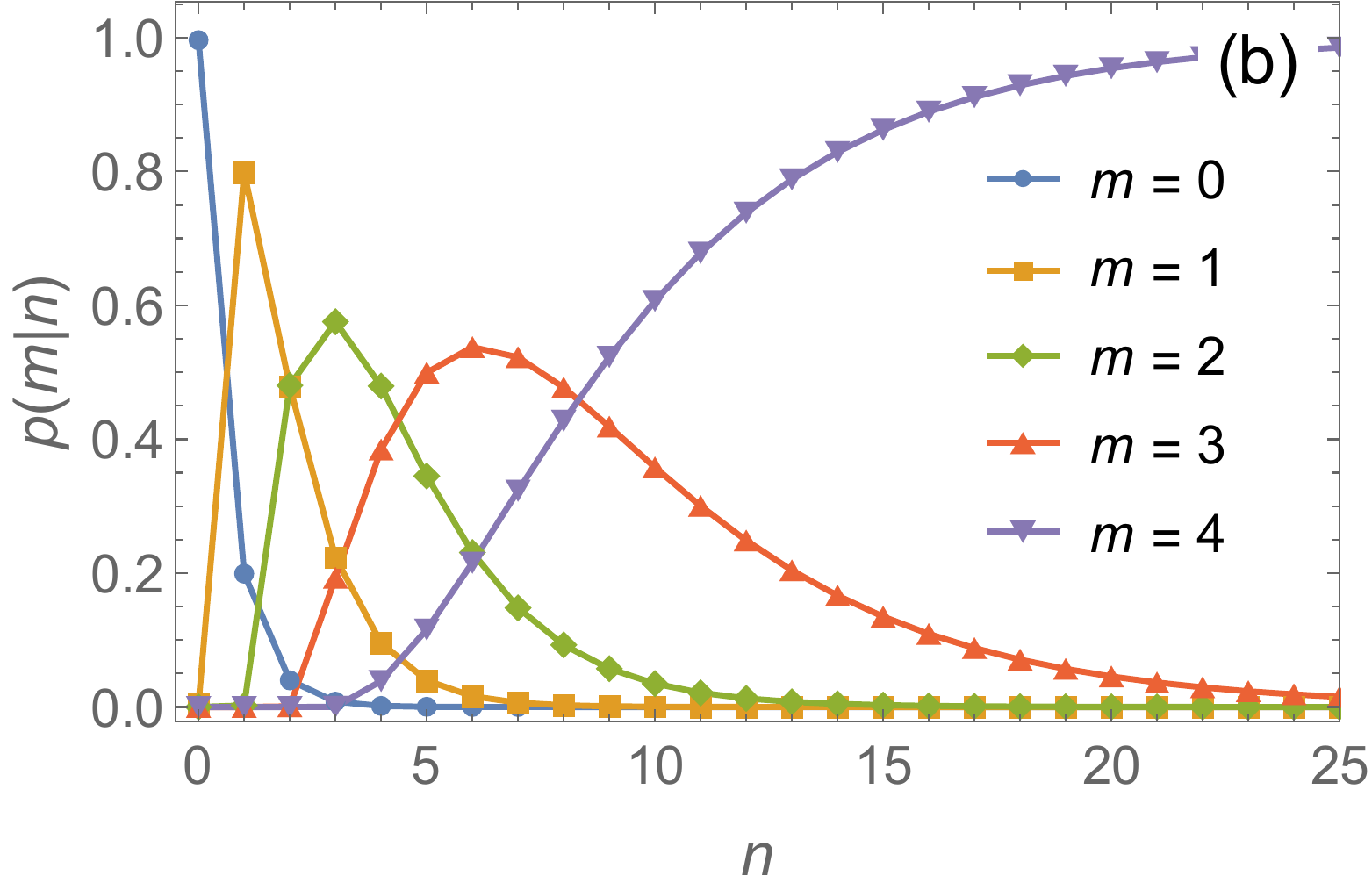}\\
\includegraphics[width=\linewidth]{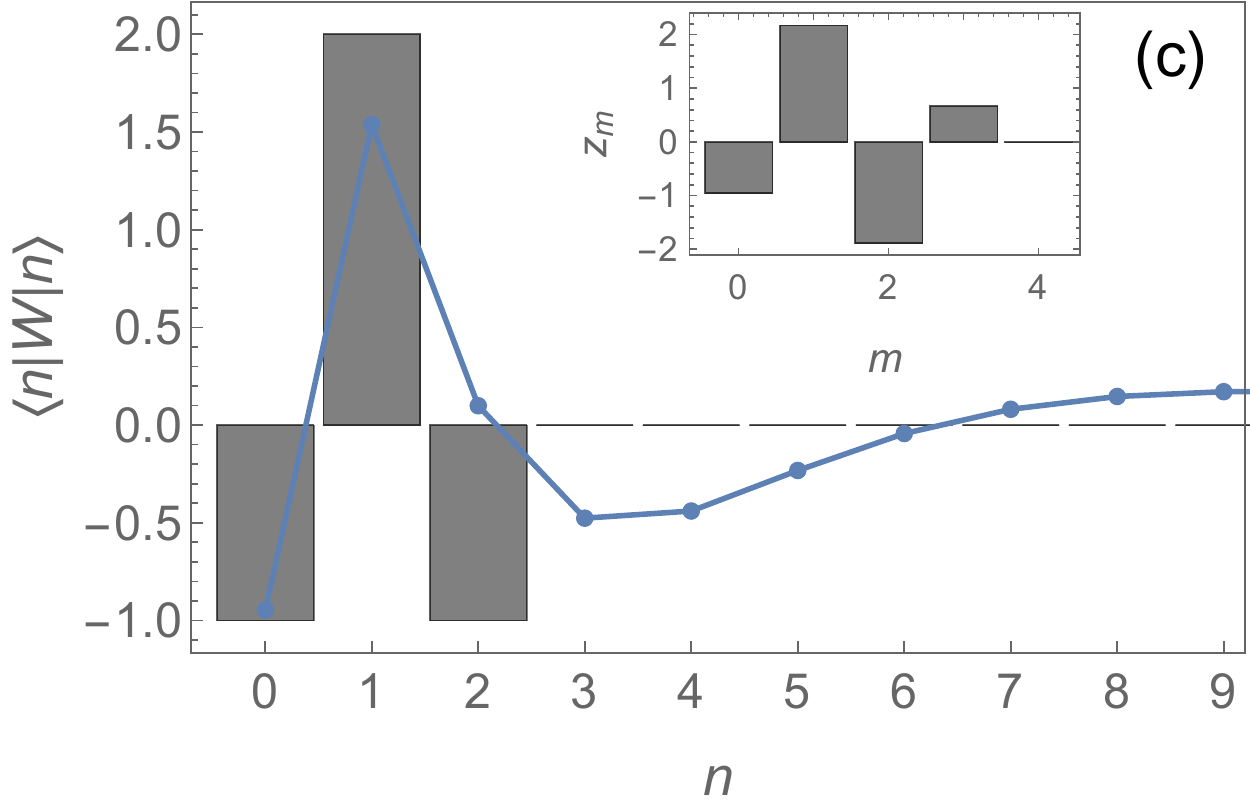}
\caption{Approximation of the single-photon non-classicality witness by the four-detector measurement setup: detection scheme (a), Fock-basis decomposition coefficients of the POVM elements $\Pi_m$ (b), and comparison of the ideal witness $W$ (gray bars) and the constructed witness $W_4$ (blue line) - plot (c). The inset in plot (c) shows the found decomposition coefficients $z_m$ in Eq.~(\ref{eqn:witness_decomposition}). The detection efficiency $\eta = 0.8$ and the dark count rate $\varepsilon = 0.001$ were used for the calculations.}
\label{fig:four-detectors}
\end{figure}

\subsection{Non-classicality witnessing under realistic measurement conditions}

The non-classicality witness $W$, described by Eq.~(\ref{eqn:witness_single_photon}), requires a photon number resolving measurement. To stay more realistic, it is worth constructing a witness, which can be measured with usual single-photon detectors, possessing final efficiency and dark count rate.

Let us consider the measurement setup, shown in Fig.~\ref{fig:four-detectors}a and consisting of 4 single-photon detectors with the detection efficiency $\eta$ and the dark count rate $\varepsilon$. The five possible outcomes of the measurement correspond to detection of $m = 0$, 1, 2, 3, and 4 counts respectively and can be described by the positive operator-valued measure (POVM) $\{\Pi_0, \ldots, \Pi_4\}$. The POVM elements have the following Fock-state basis representation:
\begin{equation}
    \label{eqn:POVM}
    \Pi_m = \sum_{n = 0}^\infty |n \rangle \langle n | p(m|n),
\end{equation}
where
\begin{multline}
    \label{eqn:p_m_n}
    p(m|n) = \sum_{k = 0}^m \frac{4!(-1)^{m - k} (1 - \varepsilon)^{4 - k}}{(4 - m)! k! (m - k)!} \\{} \times \left(1 - \frac{4 - k}{4} \eta \right)^n
\end{multline}
is the probability of detecting $m$ counts if the input state of the measurements scheme in Fig.~\ref{fig:four-detectors}a is the Fock state $n$ (Fig.~\ref{fig:four-detectors}b).

Following the ideas from Eq.~(\ref{eqn:representation}), one can try to approximate the witness operator $W$, introduced by Eq.~(\ref{eqn:witness_single_photon}), in terms of the available POVM elements:
\begin{equation}
    \label{eqn:witness_decomposition}
    W \approx W_4 \equiv \sum_{m=0}^4 z_m \Pi_m,
\end{equation}
where the coefficients $z_m$ (see the inset in Fig.~\ref{fig:four-detectors}c) can be found, for example, by minimization of the quadratic distance between $W$ and $W_4$ (the fidelity $F$ cannot be used here because neither $W$ nor $W_4$ are positive semi-definite operators):
\begin{equation}
    \label{eqn:POVM_min}
    \min _{\{z_m\}} \sum_{n=0}^\infty \left(\langle n | W | n \rangle - \langle n | W_4 | n \rangle \right)^2.
\end{equation}
Fig.~\ref{fig:four-detectors}c shows the resulting witness operator $W_4$.  While being different from the ideal witness $W$ because of detectors' non-ideality, it is still suitable for detection of non-classicality. During the performed numerical calculations, we assumed that the detection efficiency equals $\eta = 0.8$ and the dark count probability is $\varepsilon = 0.001$.

Similarly to Eq.~(\ref{eqn:witness_0}), the maximal classical value $W_{40}$ of the constructed witness corresponds to the coherent state $|\alpha \rangle$ with $\alpha = 1.176$, maximizing $W_4(|\alpha|) = \langle \alpha | W_4 |\alpha \rangle$:
\begin{equation}
    \label{eqn:witness_decomposed_0}
    W_{40} = \max_\alpha W_4(|\alpha|) = 0.248.
\end{equation}

The witness value, reached for the single-photon state, equals $\langle 1 | W_4 | 1\rangle = 1.538 > W_{40}$. Unlikely the ideal witness $W$ yielding zero variance for the state $|1\rangle$, the variance of the observable $W_4$ for the single-photon state is $1.59$. Therefore, one needs to perform at least several repetitions of the measurement to be sure that the results are incompatible with the assumption of a {non-negative-P-function} input state if the state $|1\rangle$ is supplied.

The {"classically"} emulated single-photon state, discussed in the previous sections, yields the mean value $\langle W_4 \rangle = 1.537$, which still noticeably exceeds the classical limit. The excess variance  of the observable $W_4$ is $1.8\times 10^3$. Therefore, the number of the measurement repetitions, required for reliable proof of the single-photon state non-classicality, remains approximately the same as for the measurement of the ideal witness $W$.

\begin{figure}[tp]
\includegraphics[width=\linewidth]{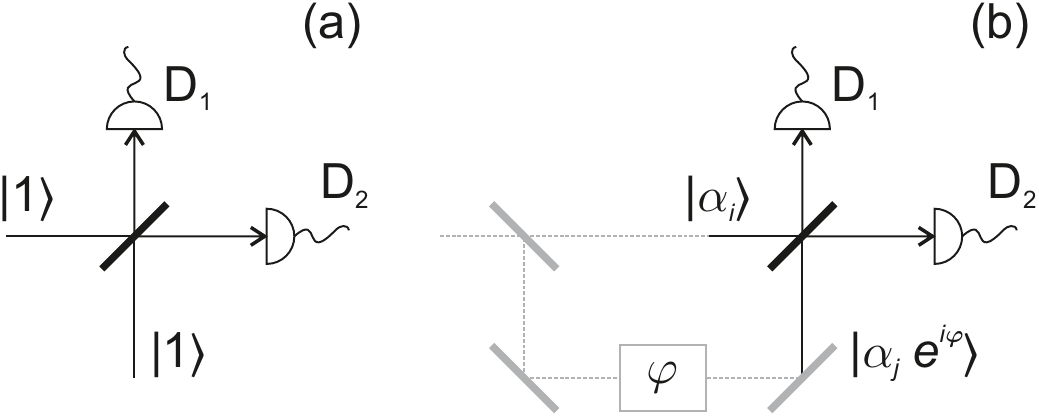}
\caption{Scheme of observing Hong-Ou-Mandel effect with single photons (a) and the setup for its {"classical emulation"} (b). Gray dashed lines show how the phase-averaged coherent states can be generated by variable splitting a reference coherent state with subsequent application of a random phase shift $\varphi$ in one of the arms.}
\label{fig:HOM_scheme}
\end{figure}

\section{Hong-Ou-Mandel effect}

Let us illustrate a  measurement emulation scheme by an archetypal quantumness demonstrator: the Hong-Ou-Mandel single-photon interference. If one has a single photon per each entry port of the 50/50 BS (Fig.~\ref{fig:HOM_scheme}a), in case of the ideal interference of both input fields $a$ and $b$, the probability $p_{12}$ of having the detectors ${D_1}$ and $D_2$ clicking  simultaneously is zero. If the interference isn't ideal (for example, due to imperfect overlapping of the pulses or misaligned polarization), $p_{12}\neq0$ and increasing with worsening of interference  \cite{PhysRevLett.113.070403,Tiedau_2018}. If the detectors have the efficiency $\eta$ and do not distinguish modes in the impinging fields, a registration of a click on $j$-th detector is described by the expression~\cite{Tiedau_2018} 
\[\Pi_j = 1-{:\exp\left\{-\frac{\eta}{2}(a^{\dagger}a+b^{\dagger}b\pm fa^{\dagger}b \pm f^*b^{\dagger}a)\right\}:},\] 
where the signs ``+'' and ``$-$'' correspond to the $j = 1$ and 2 respectively, the operators $x^{\dagger}$, $x$ are creation and annihilation operators for $x$-th mode, $x=a,b$; $::$ denotes the normal ordering, and the parameter $f$ describes the degree of the overlap. Upon considering, for simplicity, a real and positive $f$, the probability of \emph{both} detectors clicking is \[p_{12}=\langle 1_a,1_b|\Pi_1 \Pi_2 |1_a,1_b\rangle=(1-f^2)\eta^2/2,\]
where the $|1_a,1_b\rangle$ describes the single-photon Fock states in the modes $a$ and  $b$. 
 
The following scheme reproduces the Hong-Ou-Mandel effect (Fig.~\ref{fig:HOM_scheme}b): the randomly chosen phase-averaged coherent states are produced by appropriate splitting of an input coherent state, while the additional random phase shift $\varphi$ introduces the effect of the phase averaging. In this manner the representation of $\rho'$ for the input single-photon states can be built in terms of the phase-averaged coherent states,
 \begin{equation}
    \label{eqn:decomposition_11}
    \rho' = \sum_{k,l} c_k c_l\, \rho^a_k\rho^b_l.
\end{equation}
For the probe state $\rho^a_k\rho^b_l$ with a non-ideal overlap, the two-detector click probability now reads
\begin{equation}
    \label{eqn:HOM_prob11}
    p_{12}^{kl} =  \frac{1}{2 \pi} \int d\varphi \left[ 1 - p_+^{kl}(\varphi) \right]\left[ 1 - p_-^{kl}(\varphi) \right]\,,
\end{equation}
where
\[ p_\pm^{kl}(\varphi) =  \exp\left[-\frac{\eta}{2} \left(|\alpha_k|^2 + |\alpha_l|^2 \pm 2  f |\alpha_k| |\alpha_l| \cos \varphi\right) \right].\]
Let us now estimate to which extent the  representation (\ref{eqn:decomposition_11}) is more expensive in terms of the necessary number of state copies. From  Eq.~(\ref{eqn:decomposition_11}), for $\eta = 0.8$ and the overlap $f^2 = 0.95$, one gets  $p_{12}=0.017$.  For the input states being true single photons, a single-trial variance is less than unity. However, the single-trial variance, estimated according to Eq.~(\ref{eqn:noise}), is  $\operatorname{Var}(p_{12}) =1.5 \times 10^4$. So, one needs the number of measurement runs (samples of probe states), $N$, of about $10^6$  for reliable demonstration of Hong-Ou-Mandel effect. However, it is worth reminding here that to generate just a single photon, for example, by the spontaneous down-conversion, one needs about  $10^5 - 10^{12}$ pumping photons \cite{spdc}. 

\begin{figure}[tp]
\includegraphics[width=0.6\linewidth]{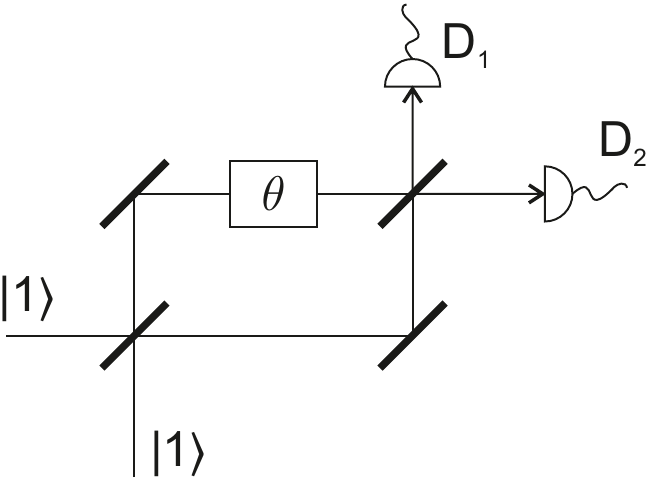}
\caption{Scheme of interferometer for phase estimation using 2-photon NOON-state.}
\label{fig:NOON_scheme}
\end{figure}

\begin{figure}[tp]
\includegraphics[width=\linewidth]{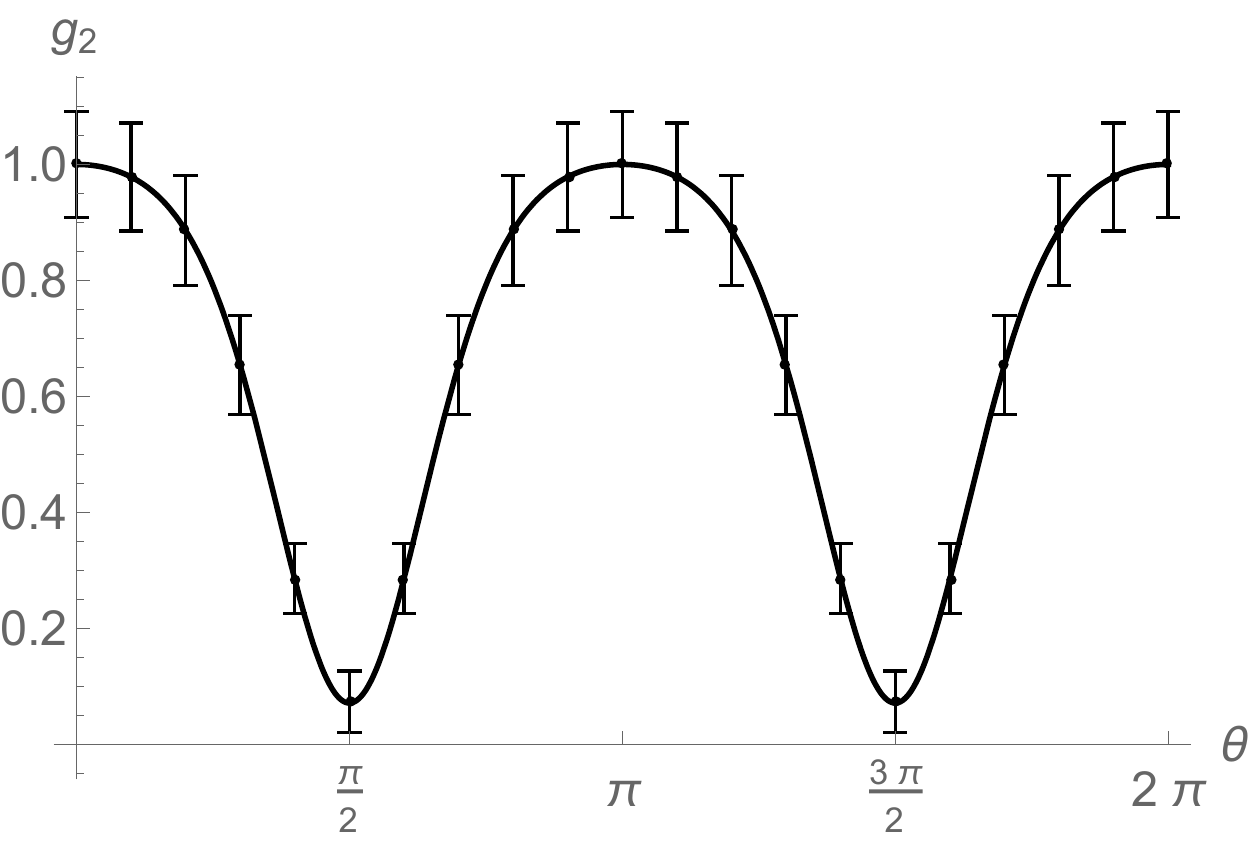}\\
\caption{Dependence of normalized coincidence rates $g_2(\theta)$ on the measured phase shift $\theta$. Solid lines indicate the dependence for interference of two single-photon states. Points and error bars show the values and the standard deviations for {"classically"} emulated single-photon states for $N = 10^8$ repetitions.}
\label{fig:NOON_g2}
\end{figure}

\section{Phase estimation with 2-photon NOON-state}

Here we show how one can emulate the phase estimation with NOON states using our  scheme. 

The state obtained after the interference of 2 photons at a beamsplitter, is the 2-photon NOON-state $(|2\rangle_a|0\rangle_b - |0\rangle_a |2\rangle_2) / \sqrt 2$ and, therefore, can be used for sensitivity enhancement in phase estimation. For the scheme, shown in Fig.~\ref{fig:NOON_scheme}, the probability of encountering both photons in one arm (coupled to either $D_1$ or $D_2$) equals
\begin{equation}
    p_2^{(0)}(\theta) = \frac{1 + f}{4} \sin^2 \theta,
\end{equation}
while the probability of having one photon in each arm is
\begin{equation}
    p_{11}^{(0)}(\theta) = \frac{1-f}{2} + \frac{1+f}{2} \cos^2 \theta,
\end{equation}
where $\theta$ is the phase shift to be measured.

Similarly to the previous section, one can calculate the probability of coincidence count $p_{11}$ and the unconditional probabilities of single-photon detection $p_{1\cdot}$ and $p_{\cdot 1}$ and introduce the normalized coincidence rate:
\begin{multline}
\label{eqn:phase_estimation_g2}
    g_2 (\theta) = \frac{p_{11}}{p_{1\cdot} p_{\cdot 1}} \\{} = \frac{16 \left(\eta  (1-\epsilon ) (z +\eta  (3-f-4 \epsilon)+8 \epsilon )+4 \epsilon
   ^2\right)}{\left(\eta (1-\epsilon ) z+\eta  (1-\epsilon ) (8-f \eta -\eta )+8 \epsilon
   \right)^2}
\end{multline}
where $z = (1 + f) \eta \cos 2 \theta$. The solid line in Fig.~\ref{fig:NOON_g2}a shows the dependence of $g_2(\theta)$ on the phase shift $\theta$.

For the 2-mode probe state $\rho_i \otimes \rho_j$, the probabilities of clicks on both detectors  are
\begin{multline}
    \label{eqn:phase_estimation_prob11}
    p_{11}(i,j; \theta) = \frac{1}{2 \pi} \int d\varphi \left[ 1 - p_-(|\alpha_i|, |\alpha_j|, \theta, \varphi) \right] \\ {} \times \left[ 1 - p_+(|\alpha_i|, |\alpha_j|, \theta, \varphi) \right],
\end{multline}
\begin{equation}
    \label{eqn:phase_estimation_prob10}
    p_{1\cdot}(i,j)  = \frac{1}{2 \pi} \int d\varphi \left[ 1 - p_-(|\alpha_i|, |\alpha_j|, \theta, \varphi) \right],
\end{equation}
and 
\begin{equation}
    \label{eqn:phase_estimation_prob01}
    p_{\cdot 1}(i,j) = \frac{1}{2 \pi} \int d\varphi \left[ 1 - p_+(|\alpha_i|, |\alpha_j|, \theta, \varphi) \right],
\end{equation}
where
\begin{multline}
    \label{eqn:phase_estimation_p0}
    p_\pm(x,y, \theta, \varphi) = (1 - \varepsilon)  \exp\Bigl[-\frac{\eta}{2} \Bigl(x^2 (1 \pm \cos \theta) \\{} + y^2 (1 \mp \cos \theta) \pm 2 \sqrt f x y \sin \theta \sin \varphi\Bigr) \Bigr].
\end{multline}

The results of the calculation of the normalized second-order functions for the true and emulated states are shown in Fig.~\ref{fig:NOON_g2}. The values $\eta = 0.8$, $\varepsilon = 0.001$, $f = 0.95$ and $N = 10^8$ repetitions were assumed.

\begin{figure}[tp]
\includegraphics[width=0.75\linewidth]{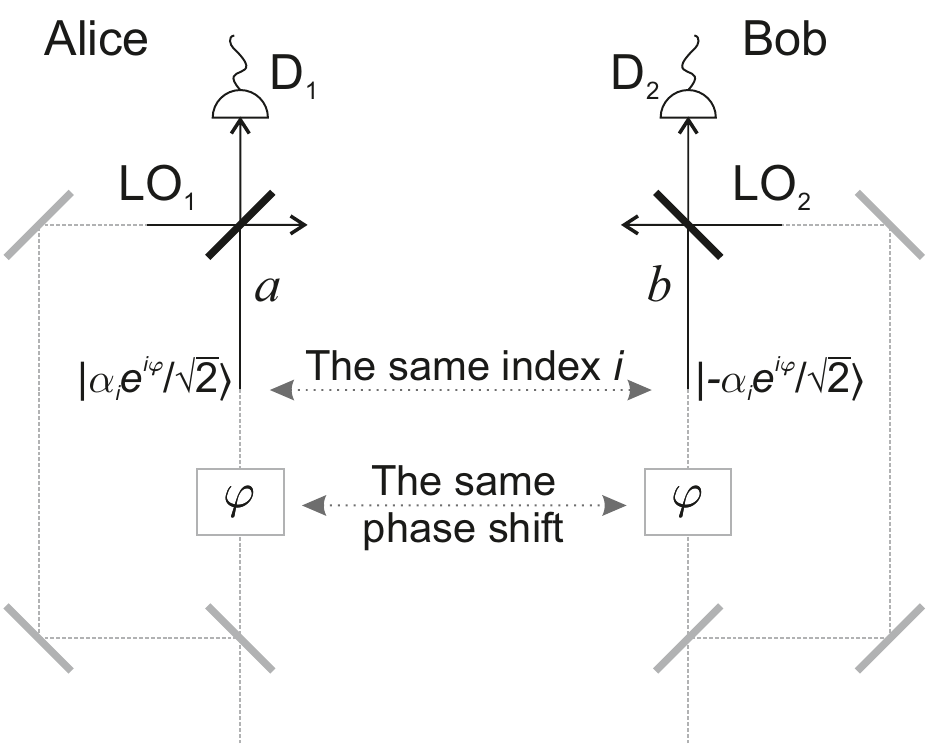}
\caption{Detection scheme for classical emulation of 2-mode entanglement and Bell-type measurements with phase-averaged probe coherent states. Alice and Bob prepare the same amplitudes and opposite phases of the coherent states.}
\label{fig:entanglement emulation}
\end{figure}

\section{Bell inequalities violation}

Another famous manifestation of quantumness is Bell-type inequalities violation for a distinguishable (for example, spatially separated) quantum systems. Let us show here how it is possible to emulate the state of two entangled modes $a$ and $b$ sharing a single photon state $|\Psi_{ab}\rangle\langle\Psi_{ab}|$, where $|\Psi_{ab}\rangle = \left(|1\rangle_a |0 \rangle_b - |0 \rangle_a |1 \rangle_b \right)/\sqrt{2}$, and to demonstrate violation of the Clauser-Horn inequality \cite{PhysRevD.10.526} using a modification of the scheme discussed in Ref.~\cite{wildfeuer2007}. In  Ref.~\cite{wildfeuer2007} the coherently displaced signals are measured with simple on-off detectors and the  following inequality is considered:
\begin{multline}
    \label{eqn:J0_definition}
    -1\le j_0 = q(\alpha, \gamma) - q(\alpha, \delta) + q(\beta, \gamma) + q(\beta, \delta)\\{} - q_a(\beta) - q_b(\gamma)\le 0,
\end{multline}
where the  single-detector no-click probabilities  and coinciding two-detector no-click probabilities to be measured for the coherently displaced input state are defined as 
\[
q_x(\mu) = \langle Q_x(\mu) \rangle =\left\langle D_x(\mu) \Pi_{\text{off}}^{(x)} D_x^\dag(\mu) \right\rangle, \]
and  
\[q(\mu, \nu) = \langle Q_a(\mu)Q_b(\nu) \rangle,\] 
where  the operator $D_x(\mu)=\exp\{\mu x^{\dagger}-\mu^*x\}$ describes coherent displacement of $x$-th mode by the amplitude $\mu$, $x=a,b$, implemented by mixing the mode with a local oscillator (LO) field at a beamsplitter (Fig.~\ref{fig:entanglement emulation}). The operator $\Pi_{\text{off}}^{(x)}$ describes  appearance of count absence on the detector measuring the $x$-th mode. For the detectors with the efficiency $\eta$ this  operator can be expressed as
\[\Pi_{\text{off}}^{(x)} = (1 - \eta)^{n_x}\] 
in terms of the number operator $n_x=x^{\dagger} x$.
The value of $j_0$ can be minimized for all possible shifts $\alpha$, $\beta$, $\gamma$, and $\delta$. To have the minimal $j_0$, one needs setting   $\alpha = -\delta = \mu_1$ and $\gamma = -\beta = \mu_2$. Particular values of $\mu_{1,2}$ depend on the efficiency. For example,  the minimal value of $j_0$ and the optimal amplitudes $\mu_1$ and $\mu_2$ for ideal detectors ($\eta=1$) are equal to ($-1.172$, 0.563, 0.165); for $\eta = 0.95$ the optimal values are ($-1.118$, 0.587, 0.177) , for $\eta = 0.90$ they are ($-1.066$, 0.615, 0.191). Taking into account the relations between the optimal amplitudes and inequality (\ref{eqn:J0_definition}), one can introduce the observable 
\begin{eqnarray}
\nonumber 
J_0=Q_a(-\mu_2)Q_b(\mu_2) - Q_a(\mu_1)Q_b(- \mu_1)-\\
\nonumber
Q_a(- \mu_2) [1 - Q_b(-\mu_1)] -[1 - Q_a(\mu_1)] Q_b(\mu_2),
\end{eqnarray} 
corresponding to the quantity $j_0 = \langle J_0 \rangle$.
To emulate $|\Psi_{ab}\rangle\langle\Psi_{ab}|$, it is convenient to represent this state  as a result of beam-splitting of a single-mode one-photon state. Emulating this single-photon state with a mixture of phase-averaged coherent states, $|\Psi_{ab}\rangle\langle\Psi_{ab}|$ can be rewritten in the form of Eq.~(\ref{eqn:representation}) with 
\begin{eqnarray}
\nonumber
\rho_j = \frac{1}{2 \pi} \int_0^{2\pi} d\varphi |f_j(\varphi)\rangle \langle f_j(\varphi)|, 
\end{eqnarray}
where the states $|f_j(\varphi)\rangle$ are products of the coherent states of the modes $a$ and $b$:   \begin{eqnarray}
\nonumber
|f_j(\varphi)\rangle={|{\alpha_j} e^{i \varphi}/{\sqrt 2}\rangle_a}{|-{\alpha_j} e^{i \varphi}/{\sqrt 2}\rangle_b}.
\end{eqnarray}
Factorization of the probe states implies that, in contrast to the emulated splitting of a single-photon state, they can be generated separately by Alice and Bob.

A scheme, suitable for implementation of the discussed emulation of the state $|\Psi_{ab}\rangle\langle\Psi_{ab}|$ is shown in Fig.~\ref{fig:entanglement emulation}. For each trial, the two parties choose the same random index $j$ of the probe state and the same random phase shift $\varphi$ and prepare the coherent states $|\alpha_j e^{i\varphi} / \sqrt 2 \rangle_a$ and $|-\alpha_j e^{i\varphi} / \sqrt 2 \rangle_b$ of the modes $a$ and $b$ respectively. Input BS's in Fig.~\ref{fig:entanglement emulation} allows one to realize required coherent shifts. 
The expectation values of $J_0$ over such probe states can be found using the rules
\[\operatorname{Tr}\left\{\rho_j Q_a(\mu)Q_b(\nu) \right\} =\, \frac{1}{2 \pi} \int d\varphi q_j(\mu, \varphi) q_j(-\nu, \varphi),\]
and
 \[\operatorname{Tr}\left\{\rho_j Q_{a(b)}(\mu) \right\}  =\, \frac{1}{2 \pi} \int d\varphi q_j(\pm\mu, \varphi),\]
where $q_j(\nu, \varphi) = \exp(- \eta| {\alpha_j e^{i\varphi}}/{\sqrt 2} - \nu|^2)$. Since the Alice's and Bob's parts of the probe states interfere with their LO fields only, the phase stability between Alice's and Bob's coherent-state sources is not required.

As it is to be expected, the mean value of the measured observable remains approximately the same as for the true state, $\langle J_0\rangle = -1.118,-1.077$ for $\eta = 0.95,0.9$,  while the single-trial variances are expectedly large: $\operatorname{Var}(J_0) = 5.1 \times 10^3$ and $5.0 \times 10^3$ respectively. The numbers of the measurement repetitions, required for reliable non-classicality demonstration, are $N \gtrsim 1.5 \times 10^6$ for $\eta = 0.95$ and $N \gtrsim 3.4 \times 10^6$ for $\eta = 0.90$.

\section{Conclusions}

We have shown that results of quantum measurements can be emulated using only {quantum states with non-negative P functions}. For that purpose, one just needs to know how the quantum state can be prepared with a set of {non-negative-P-function} states by binary labeling each probe and using that classical information during the measurement. The price for the ability to use {such ``classical''} light sources is the necessity to tailor our method for each particular state, and  larger number of measurement runs required for getting reliable results. Generally, this number might become so large  as to render our emulation procedure unfeasible. However, we have demonstrated that for some important and interesting states and measurement schemes our approach is feasible and can be even more efficient in terms of {``classical''} resources necessary to generate and measure the state. The proposed approach is likely to be a handy toolbox for proof-of-principle experiments for testing quantum effects; for verification of proper functioning of measurement setups for fundamental quantum experiments before the required non-classical states become available; and for design of affordable demonstration breadboards for education and science dissemination. The presented procedure provides a systematic tool for ``quantum-inspired'' metrology by translating quantum measurement protocols into their classical counterparts, where the {non-negative-P-function}-state signals are combined in the optimal way in order to enhance the sensitivity. 

\begin{acknowledgments} 
	A. M. and D. M. gratefully acknowledge support from The Belarusian Republican Foundation for Fundamental Research (F21KOR-002,F21IKR-003), the EU project PhoG 820365 and the NATO project NATO SPS - G5860. Y. S. T. and H. J. acknowledge support by the National Research Foundation of Korea (NRF) (Grant Nos. NRF-2019R1A6A1A10073437, NRF2019M3E4A1080074, NRF-2020R1A2C1008609 and 2020K2A9A1A06102946) via the Institute of Applied Physics at Seoul National University and by Institute of Information \& communications Technology Planning \& Evaluation (IITP) grant funded by the Korea government (MSIT) (Nos. 2021-0-01059 and 2021-2020-0-01606).
\end{acknowledgments}

\appendix

\section*{Appendix: Developing classical representations}

\subsection{{Representation fidelity and systematic error of emulation}}

Let us consider emulation of measuring an observable $A$ in a quantum state $\rho_\text{true}$ by using the representation 
\[\rho_\text{true} \approx \rho = \sum_i c_i \rho_i\] 
in terms of semi-classical probe states $\{\rho_i\}$. The systematic error of the emulation equals 
\begin{equation}
    \label{eq:systematic_error}
    \delta_\text{sys.} = |\langle A \rangle _\rho - \langle A \rangle _\text{true}| = |\mathrm{Tr}\{A(\rho-\rho_{\mathrm{true}})\}|.
\end{equation}
If the exact operator $A$, describing the measured observable, is known when the emulation protocol is designed, Eq.~(\ref{eq:systematic_error}) can be used directly for assessment of the particular representation quality, since it defines the distance between $\rho$ and $\rho_\text{true}$ appropriate for the particular measurement. Otherwise, it is instructive to connect the upper bound of the systematic error for an \textit{arbitrary} observable $A$, satisfying certain reasonable constraints, with the representation fidelity
\begin{equation}
    F(\rho, \rho_\text{true}) = \Bigl[ \operatorname{Tr} \sqrt{\sqrt{\rho}\rho_\text{true}\sqrt{\rho}} \Bigr]^2.
\end{equation}

Measurement of an observable $A$ is characterized {by a set of projective operators $E_i$, forming} a positive operator valued measure (POVM) $\{E_i\}$ with $i$-th outcome being mapped to the eigenvalue $\lambda_i$ of the operator $A$:
\begin{equation}
    \langle A \rangle _\rho = \sum_i \lambda_i p_i, \quad p_i= \operatorname{Tr}( E_i \rho),
\end{equation} 
\begin{equation} \langle A \rangle _\text{true} = \sum_i \lambda_i q_i, \quad q_i = \operatorname{Tr}( E_i \rho_\text{true}),
\end{equation}
and
\begin{equation}
    \delta_\text{sys.} = \Bigl| \sum_i \lambda_i (p_i - q_i) \Bigr|.
\end{equation}

The fidelity of the representation is connected to distinguishability of the states $\rho$ and $\rho_\text{true}$ by the optimal POVM measurement \cite{fuchs1995mathematical}, which is at least as sensitive as the particular measurement {characterized by the POVM $\{E_i\}$ and associated with} the observable $A$:
\begin{multline}
    \label{eq:fidelity_by_POVM}
    F(\rho, \rho_\text{true}) = \min\limits _{\text{POVM } \{E'_i\}} F_\text{B}^2(\{\operatorname{Tr}(E'_i \rho)\},\{\operatorname{Tr}(E'_i \rho_\text{true})\}) \\ \le F_\text{B}^2( p, q),
\end{multline}
where $F_\text{B}(u, v) = \sum_i \sqrt{u_i v_i}$ is the Bhattacharyya coefficient of classical probability distributions $u$ and $v$; $p = \{p_i\}$ and $q = \{q_i\}$ are the probability distributions for the measurement of $A$ for $\rho$ and $\rho_\text{true}$.

The right-hand side of Eq.(\ref{eq:fidelity_by_POVM}) depends on the particular choice of the measured observable $A$. To get a universal bound, one can maximize $F_\text{B}(p,q)$ over appropriate measurements:
\begin{equation}
    \label{eq:fidelity_by_max_FB}
    F(\rho, \rho_\text{true}) \le \max\limits_{p, q, \lambda} F_\text{B}^2( p, q), 
\end{equation}
conditioned by 
\begin{equation}
    \label{eq:constraints}
     \sum_i p_i = 1, \quad \sum_i q_i = 1, \quad \Bigl| \sum_i \lambda_i (p_ i - q_i) \Bigr| = \delta_\text{sys.}
\end{equation}

Let us assume that any considered observable $A$ is known \textit{a priori} to have limited eigenvalues: $|\lambda_i| \le M$. For example, if the probability of detecting a photon or a coincidence count is measured, we have $M = 1$. 

Optimization of Eq.~(\ref{eq:fidelity_by_max_FB}) over $\lambda$ immediately shows that the choice 
\[\lambda_i = M \sgn {p_i - q_i} \] 
is optimal for maximization of $F_\text{B}(p,q)$. The last constraint in Eq.~(\ref{eq:constraints}) is transformed into 
\[\sum_i |p_i - q_i| = \delta_\text{sys.} / M.\] 
Pairwise variations of $(p_i,q_i)$ and $(p_j,q_j)$ for all such $i$ and $j$ that 
$\sgn{p_i - q_i} = \sgn{p_j - q_j}$ lead to the optimality condition $p_i : q_i = p_j : q_j$. Finally, one arrives at the condition 
\begin{equation}
    \sum_{i: \; p_i > q_i} (p_i + q_i) = \sum_{i: \; p_i < q_i} (p_i + q_i) = 1
\end{equation}
yielding the maximal value
\begin{equation}
    \label{eq:FB_max}
    \max\limits_{p, q, \lambda} F_\text{B}^2( p, q) = 1 - \delta_\text{sys.}^2 / (2M)^2.
\end{equation}

Combining Eqs.~(\ref{eq:fidelity_by_max_FB}) and (\ref{eq:FB_max}) and solving for $\delta_\text{sys.}$, we arrive at the following upper bound for the systematic error of emulation:
\begin{equation}
    \delta_\text{sys.} \le 2 M \sqrt{1 - F(\rho, \rho_\text{true})}.
\end{equation}

\subsection{Rate of convergence of $\left<A\right>_N$}

Suppose that the set of {"classical"} component probe states $\{\rho_j\}$ are used to emulate the target true state $\rho_\text{true}\approx\sum_jc_j\rho_j$. In the perspective of statistical inference, one can rewrite $\left<A\right>_N$ defined in Eq.~(3) of the main text as
\begin{equation}
\left<A\right>_N=\dfrac{\zeta}{N}\sum_j\sum_{l}\sgn{c_j}a_ln_{jl}\,,
\end{equation} 
where 
\[\zeta=\sum_j|c_j| = \zeta_+ + \zeta_-,\] 
and now $\left<A\right>_N$ is a double sum over all probe states employed and eigenvalues $a_l$ of the observable $A$ measured. The frequencies $n_{jl}$ of having $l$-th result of the measurement with $j$-th probe are summed to the total number of trials, $N$:  $\sum_{j,l}n_{jl}=N$. Thus the relative frequencies $\nu_{jl}=n_{jl}/N$ inherently follow a multinomial distribution with respect to the indices $j$ and $l$, with the statistical average
\begin{equation}
\overline{\nu_{jl}\nu_{j'l'}}=\begin{cases}
[\delta_{l,l'}\widetilde{p}_{jl}+(N-1)\widetilde{p}_{jl}\widetilde{p}_{jl'}]/N & \text{when }j=j'\,,\\
(N-1)\widetilde{p}_{jl}\widetilde{p}_{j'l'}/N & \text{otherwise}\,.\\
\end{cases}
\end{equation} 
determined by the observation $\overline{\nu_{jl}}=\widetilde{p}_{jl}=p_jp_{jl}$, $p_j=|c_j|/\zeta$ and $p_{jl}=\langle{a_l}|{\rho_j}|{a_l}\rangle$. These immediately give
\begin{align}
&\,\overline{\left(\left<A\right>_N-\left<A\right>\right)^2}\nonumber\\
=&\,\sum_{j,j',l,l'}\sgn{c_j}\sgn{c_{j'}}a_la_{l'}\overline{(\nu_{jl}\nu_{j'l'}-\widetilde{p}_{jl}\widetilde{p}_{j'l'})}\nonumber\\
=&\,\dfrac{\zeta^2}{N}\left[\sum_jp_j\Tr{\rho_j A^2}-\left(\sum_jp_j\sgn{c_j}\Tr{\rho_j A}\right)^2\right]\nonumber\\
=&\,\dfrac{1}{N}\left[\zeta\,\Tr{(\zeta_+\rho_++\zeta_-\rho_-) A^2}-\left<A\right>^2\right]=\dfrac{\Delta_{AB}}{N}\,.
\end{align}
Alternatively, one arrives at this result by considering the extended model [$\left<A\right>=\Tr{(A\otimes B)\rho_\textsc{c}}$] discussed in the text and recognizing the fact that each \emph{independently} sampled eigenvalue incurs a quantum variance of $\Delta_{AB}$, such that scaling $\Delta_{AB}$ with $N$ gives the right answer. Hence, in the limit of large $N$, we indeed expect that \[\left<A\right>-\left<A\right>_N=O(1/\sqrt{N}).\]

On the other hand, for the same number of copies $N$, if one can generate $\rho_\text{true}$ directly, then the naive linear estimator $\left<A\right>_\textsc{lin}=\sum_la_l\nu_l$ for $\left<A\right>$, where $\nu_l=n_l/N\rightarrow p_l=\langle{a_l}|{\rho_\text{true}}|{a_l}\rangle$, leads to
\begin{equation}
\overline{\left(\left<A\right>_\textsc{lin}-\left<A\right>\right)^2}=\dfrac{\Delta_A}{N}
\end{equation} 
owing to the basic multinomial rule 
\[\overline{(\nu_j - p_j)(\nu_k-p_k)}=\frac{1}{N}(p_j\delta_{j,k}-p_jp_k).\] 
As argued in the text, the fact that $\Delta_{AB}>\Delta_{A}$ simply reiterates that non-classical state emulation using {"classical"} component states reduces the complexity of quantum-state generation at the expense of a larger $N$ to estimate $\left<A\right>$ up to some fixed accuracy.

\subsection{Sampling accuracy in non-classical quantum-state emulation}

Let us show how well sampling of the probe states from the set with the help of $\rho_\textsc{c}$ (equivalent to using a classical random-number generator) can approximate the state $\rho$ from Eq.~(1) of the main text.  

After a multinomial sampling of $N_\text{s}$ copies of $\rho$ (not to be confused with $N$, the total number of copies used to estimate $\left<A\right>_N$), one obtains the estimator  $\widehat{\rho}=\zeta\sum_j\nu_j \sgn{c_j}\rho_j$, where the relative frequencies $\nu_j$ of the probe states $\rho_j$ tend to the probabilities $p_j=|c_j|/\zeta$ for $N_\text{s}\gg1$. We may consider the mean squared-error (MSE), $\mathrm{MSE}= \overline{\Tr{(\widehat{\rho}-\rho)^2}}$ with $\overline{\,\vphantom{\operatorname{Tr}{(\widehat{\rho}-\rho)^2}}\cdot\,}$ denoting the statistical mean (expectation value), as the figure of merit for determining the accuracy of such a sampling with a given value of $N_\text{s}$ with respect to the actual state $\rho=\sum_jc_j\rho_j$ being classically emulated. For a multinomial distribution, as 
$\overline{\nu_j} = p_j$ and \[\overline{(\nu_j - p_j)(\nu_k-p_k)}=\frac{1}{N_\text{s}}(p_j\delta_{j,k}-p_jp_k),\] 
the MSE can be easily computed to be 
\begin{equation}
	\mathrm{MSE}=\frac{1}{N_{\text{s}}}\sum_{j,j'}\!\left(|c_j||c_{j'}|\Tr{\rho_j^2}-c_jc_{j'}\Tr{\rho_j\rho_{j'}}\right).
	\label{eq:MSE_mult}
\end{equation}

An important special case corresponds to pure probe states $(\Tr{\rho_j^2} = 1)$. They allow to see more clearly into the essence of statistical noise introduced by {"classical"} emulation of quantum states. For mixed probe states $\rho_j$, their intrinsic classical noise is  masked by the assumption about their noise-less sampling. For pure probes, one can represent the MSE (\ref{eq:MSE_mult}) as
\begin{equation}
	\mathrm{MSE}=\frac{1}{N_\text{s}}\left\{\left( 1 - \Tr{\rho^2}\right) + \left( \zeta^2 -1  \right) \right\}.
	\label{eq:MSE_phys}
\end{equation}
The first term corresponds to the internal classical noise of the mixed state $\rho$, while the latter one describes the additional sampling noise introduced by classical representation of a non-classical state. If the state $\rho$ is "classical", one can find its representation with positive weights $c_j > 0$. Therefore, according to standard normalization of the density matrices $\rho$ and $\rho_j$, we have $\zeta = \sum_j |c_j| = \sum_j c_j = 1$ and the second term in Eq. (\ref{eq:MSE_phys}) vanishes. Notice that this term also vanishes when one samples the actual physical state (2) of the main text used for reproducing measurement results instead of $\rho$ having
$\mathrm{MSE}=\left( 1 - \Tr{\rho_c^2}\right)/{N_\text{s}}$. 

It is also worth noting that if $\rho_\text{true}$ is also pure, the MSE in~(\ref{eq:MSE_phys}) is defined by accuracy and purity of the representation,  and goes to zero with the fidelity going to unity.

\subsection{Optimizing the representation}

Our task is to simulate the result of quantum-state measurement.  So, the optimization task for developing the representation would consist of choosing the minimal possible number of {"classical"} probes providing for the least error is estimating a specified observable. Also, these probes themselves should be specified. The choice of the probes for the quantum state reconstruction was discussed in a number of works \cite{Mogilevtsev_2013,BayesianPRA2015,reut2017}. However, measurement-oriented optimization as discussed in the current contribution, was not carried on. We leave this discussion for the future works. Here we consider a particular case of optimization which allows shedding some light on the best choice of the probe states for the experiments.  

First, let us search for the optimal decomposition of the state $\rho_\text{true}$, which simultaneously minimizes the additional sampling noise $(\zeta^2 - 1) / N_\text{s}$ (see Eq. (\ref{eq:MSE_phys})) and the decomposition error $\Tr{(\rho - \rho_\text{true})^2}$:
\begin{equation}
    \min\limits_{\{\rho_j\},\{c_j\}} D,\quad D =   \left( \frac{\zeta^2 - 1}{N_\text{s}} + \operatorname{Tr}\{(\rho - \rho_\text{true})^2\} \right).
\end{equation}

A variation in $\rho_j$ and $c_j$ gives
\begin{multline}
\label{eq:variation}
    \delta D = 2 \sum_j \left(\frac{\zeta \sgn{c_j}}{N_\text{s}} + \operatorname{Tr}\{ \rho_j (\rho - \rho_\text{true})\} \right) \delta c_j \\ + 2 \sum_j c_j \operatorname{Tr}\{ \delta \rho_j (\rho - \rho_\text{true})\}.
\end{multline}
The normalization of density operators imposes the constraints $\sum_j \delta c_j = 0$ and $\Tr{\delta \rho_j} = 0$ on the variations. Strictly speaking, one should also impose certain constraints ensuring semi-positivity of $\rho$ during the considered variation. However, if the state $\rho$ is mixed with all its eigenvalues being strictly positive, the semi-positivity condition is not violated for any infinitely small variation.

First, let us assume that we do not require the states $\rho_j$ to be classical, i.e. we do not impose any additional constraints on $\delta \rho_j$. In that case, the variation in $\rho_j$ (the lower line in Eq. (\ref{eq:variation})) implies that the optimal representation should be accurate: $\rho = \rho_\text{true}$. Then, variation in $c_j$ leads to the equation $\sum_j \sgn{c_j}\, \delta c_j = 0$ and, finally, to positivity of all the weights: $c_j > 0$ for all $j$. Therefore, if we are not limited in terms of the structure of $\rho_j$, the optimal choice is either $\rho_\text{true}$ itself or any of its representations in the form of a positive-weight mixture (if $\rho_\text{true}$ is mixed).

The latter conclusion about the positivity of optimal $c_j$ as soon as the representation is accurate ($\rho = \rho_\text{true}$) stems from the variation in $c_j$ and remains valid regardless of any constraints imposed on $\delta \rho_j$. If, for the available set of probe states $\rho_j$, an accurate decomposition of $\rho_\text{true}$ requires negative weights, such a decomposition is not optimal. For usage of a simpler representation with a smaller number of components $\rho_j$, the advantage of having smaller sampling noise will exceedingly compensate for the loss in representation accuracy. 

Let us now assume that the set of probe states is limited by coherent states only: $\rho_j = |\alpha_j \rangle \langle \alpha_j |$. Their variations can be written as
\begin{equation}
    \delta \rho_j = (a^\dag - \alpha_j^\ast)|\alpha_j \rangle \langle \alpha_j | \delta \alpha_j + |\alpha_j \rangle \langle \alpha_j | (a - \alpha_j) \delta \alpha_j^\ast.
\end{equation}
Independence of the variations of the coherent states amplitudes implies that the following condition should be satisfied for the optimal representation:
\begin{equation}
    \langle \alpha_j |[a, \rho - \rho_\text{true}] |\alpha_j\rangle = 0.
\end{equation}
From this equation it follows, for example, that for the set of coherent probes not limited to just a single state,  difference between the optimally represented $\rho$ and $\rho_\text{true}$ cannot be proportional to a coherent or a number state. Generally, the operator  $[a, \rho - \rho_\text{true}]$ should be expressible as a sum of linearly independent operators. The number of these operators should be at least that of the probe states. 

Now let us consider a measurement-oriented version of the procedure described above. We look for the optimal component set that minimizes sampling noise  \eqref{eq:MSE_mult} for a given $N_{\text{s}}$ and under the condition that a certain measurement result is to be obtained. As the properties of the optimal components $\rho_j$ are of interest here, we shall focus only on their variation. Furthermore, we suppose that there is some desirable physical property about the decomposition $\rho$ that needs to be fixed in the form of an expectation-value constraint, namely $\Tr{{\rho}\,O}=\mu$ for some observable $O$. The corresponding distance to be minimized, $D'=N_\text{s}\mathrm{MSE}-\lambda(\Tr{{\rho}\,O}-\mu)$. This distance is parametrized by a Lagrange scalar $\lambda$. A variation in $\rho_j$ therefore gives
\begin{align}
 	\delta D'=&\,2\zeta\sum_j|c_j|\Tr{\rho_j\delta\rho_j}-2\sum_jc_j\Tr{\rho\delta\rho_j}\nonumber\\
 	&\,-\lambda\sum_jc_j\Tr{O\delta\rho_j}.
\end{align}
Since the $\rho_j$'s are quantum states, they take the form $\rho_j=W_j^\dag W_j/\Tr{W_j^\dag W_j}$. This implies the variation
\begin{equation}
 	\delta\rho_j=\dfrac{\delta W^\dag_j W_j+W^\dag_j\delta W_j}{\Tr{W_j^\dag W_j}}-\rho_j\,\dfrac{\Tr{\delta W^\dag_j W_j+W^\dag_j\delta W_j}}{\Tr{W_j^\dag W_j}}
\end{equation}
that leads to the extremal equation
\begin{align}
 	c_j\rho_j M=&\,2\zeta|c_j|(\rho_j^2-\rho_j\Tr{\rho_j^2})+c_j\rho_j\Tr{\rho_j M}\,,\nonumber\\
 	M=&\,2\rho+\lambda\,O,
	\label{eq:ext}
\end{align}
when $\delta D'$ is set to zero. 

In order to satisfy~\eqref{eq:ext}, we now need $\rho_j$ to commute with $2\rho+\lambda\,O$. This practically means that $\rho_j$ shares common eigenstates with $\rho$ and $O$. A specific situation is when $O=a^\dag a$ and $\mu$ is the mean photon number of the system. Then, an extremal set of $\rho_j$'s is some set of Fock-state mixture ($\rho_j=\sum_n |n\rangle w_{jn}\langle n|$ with $\sum_nw_{jn}=1$), which is compatible with a $\rho$ that is also a Fock state. 


\begin{thebibliography}{30}%
\makeatletter
\providecommand \@ifxundefined [1]{%
 \@ifx{#1\undefined}
}%
\providecommand \@ifnum [1]{%
 \ifnum #1\expandafter \@firstoftwo
 \else \expandafter \@secondoftwo
 \fi
}%
\providecommand \@ifx [1]{%
 \ifx #1\expandafter \@firstoftwo
 \else \expandafter \@secondoftwo
 \fi
}%
\providecommand \natexlab [1]{#1}%
\providecommand \enquote  [1]{``#1''}%
\providecommand \bibnamefont  [1]{#1}%
\providecommand \bibfnamefont [1]{#1}%
\providecommand \citenamefont [1]{#1}%
\providecommand \href@noop [0]{\@secondoftwo}%
\providecommand \href [0]{\begingroup \@sanitize@url \@href}%
\providecommand \@href[1]{\@@startlink{#1}\@@href}%
\providecommand \@@href[1]{\endgroup#1\@@endlink}%
\providecommand \@sanitize@url [0]{\catcode `\\12\catcode `\$12\catcode
  `\&12\catcode `\#12\catcode `\^12\catcode `\_12\catcode `\%12\relax}%
\providecommand \@@startlink[1]{}%
\providecommand \@@endlink[0]{}%
\providecommand \url  [0]{\begingroup\@sanitize@url \@url }%
\providecommand \@url [1]{\endgroup\@href {#1}{\urlprefix }}%
\providecommand \urlprefix  [0]{URL }%
\providecommand \Eprint [0]{\href }%
\providecommand \doibase [0]{https://doi.org/}%
\providecommand \selectlanguage [0]{\@gobble}%
\providecommand \bibinfo  [0]{\@secondoftwo}%
\providecommand \bibfield  [0]{\@secondoftwo}%
\providecommand \translation [1]{[#1]}%
\providecommand \BibitemOpen [0]{}%
\providecommand \bibitemStop [0]{}%
\providecommand \bibitemNoStop [0]{.\EOS\space}%
\providecommand \EOS [0]{\spacefactor3000\relax}%
\providecommand \BibitemShut  [1]{\csname bibitem#1\endcsname}%
\let\auto@bib@innerbib\@empty
\bibitem [{\citenamefont {Hong}\ \emph {et~al.}(1987)\citenamefont {Hong},
  \citenamefont {Ou},\ and\ \citenamefont {Mandel}}]{PhysRevLett.59.2044}%
  \BibitemOpen
  \bibfield  {author} {\bibinfo {author} {\bibfnamefont {C.~K.}\ \bibnamefont
  {Hong}}, \bibinfo {author} {\bibfnamefont {Z.~Y.}\ \bibnamefont {Ou}},\ and\
  \bibinfo {author} {\bibfnamefont {L.}~\bibnamefont {Mandel}},\ }\bibfield
  {title} {\bibinfo {title} {Measurement of subpicosecond time intervals
  between two photons by interference},\ }\href@noop {} {\bibfield  {journal}
  {\bibinfo  {journal} {Phys. Rev. Lett.}\ }\textbf {\bibinfo {volume} {59}},\
  \bibinfo {pages} {2044} (\bibinfo {year} {1987})}\BibitemShut {NoStop}%
\bibitem [{\citenamefont {Davidovich}(1996)}]{RevModPhys.68.127}%
  \BibitemOpen
  \bibfield  {author} {\bibinfo {author} {\bibfnamefont {L.}~\bibnamefont
  {Davidovich}},\ }\bibfield  {title} {\bibinfo {title} {Sub-{P}oissonian
  processes in quantum optics},\ }\href@noop {} {\bibfield  {journal} {\bibinfo
   {journal} {Rev. Mod. Phys.}\ }\textbf {\bibinfo {volume} {68}},\ \bibinfo
  {pages} {127} (\bibinfo {year} {1996})}\BibitemShut {NoStop}%
\bibitem [{\citenamefont {Giovannetti}\ \emph {et~al.}(2011)\citenamefont
  {Giovannetti}, \citenamefont {Lloyd},\ and\ \citenamefont {Maccone}}]{metro}%
  \BibitemOpen
  \bibfield  {author} {\bibinfo {author} {\bibfnamefont {V.}~\bibnamefont
  {Giovannetti}}, \bibinfo {author} {\bibfnamefont {S.}~\bibnamefont {Lloyd}},\
  and\ \bibinfo {author} {\bibfnamefont {L.}~\bibnamefont {Maccone}},\
  }\bibfield  {title} {\bibinfo {title} {Advances in quantum metrology},\
  }\href@noop {} {\bibfield  {journal} {\bibinfo  {journal} {Nature Photon.}\
  }\textbf {\bibinfo {volume} {5}},\ \bibinfo {pages} {222} (\bibinfo {year}
  {2011})}\BibitemShut {NoStop}%
\bibitem [{\citenamefont {Bell}(1964)}]{PhysicsPhysiqueFizika.1.195}%
  \BibitemOpen
  \bibfield  {author} {\bibinfo {author} {\bibfnamefont {J.~S.}\ \bibnamefont
  {Bell}},\ }\bibfield  {title} {\bibinfo {title} {On the {E}instein {P}odolsky
  {R}osen paradox},\ }\href@noop {} {\bibfield  {journal} {\bibinfo  {journal}
  {Physics Physique Fizika}\ }\textbf {\bibinfo {volume} {1}},\ \bibinfo
  {pages} {195} (\bibinfo {year} {1964})}\BibitemShut {NoStop}%
\bibitem [{\citenamefont {Glauber}(1963{\natexlab{a}})}]{PhysRev.130.2529}%
  \BibitemOpen
  \bibfield  {author} {\bibinfo {author} {\bibfnamefont {R.~J.}\ \bibnamefont
  {Glauber}},\ }\bibfield  {title} {\bibinfo {title} {The quantum theory of
  optical coherence},\ }\href {https://doi.org/10.1103/PhysRev.130.2529}
  {\bibfield  {journal} {\bibinfo  {journal} {Phys. Rev.}\ }\textbf {\bibinfo
  {volume} {130}},\ \bibinfo {pages} {2529} (\bibinfo {year}
  {1963}{\natexlab{a}})}\BibitemShut {NoStop}%
\bibitem [{\citenamefont {Sudarshan}(1963)}]{PhysRevLett.10.277}%
  \BibitemOpen
  \bibfield  {author} {\bibinfo {author} {\bibfnamefont {E.~C.~G.}\
  \bibnamefont {Sudarshan}},\ }\bibfield  {title} {\bibinfo {title}
  {Equivalence of semiclassical and quantum mechanical descriptions of
  statistical light beams},\ }\href@noop {} {\bibfield  {journal} {\bibinfo
  {journal} {Phys. Rev. Lett.}\ }\textbf {\bibinfo {volume} {10}},\ \bibinfo
  {pages} {277} (\bibinfo {year} {1963})}\BibitemShut {NoStop}%
\bibitem [{\citenamefont {Titulaer}\ and\ \citenamefont
  {Glauber}(1965)}]{PhysRev.140.B676}%
  \BibitemOpen
  \bibfield  {author} {\bibinfo {author} {\bibfnamefont {U.~M.}\ \bibnamefont
  {Titulaer}}\ and\ \bibinfo {author} {\bibfnamefont {R.~J.}\ \bibnamefont
  {Glauber}},\ }\bibfield  {title} {\bibinfo {title} {Correlation functions for
  coherent fields},\ }\href {https://doi.org/10.1103/PhysRev.140.B676}
  {\bibfield  {journal} {\bibinfo  {journal} {Phys. Rev.}\ }\textbf {\bibinfo
  {volume} {140}},\ \bibinfo {pages} {B676} (\bibinfo {year}
  {1965})}\BibitemShut {NoStop}%
\bibitem [{\citenamefont {Qian}\ \emph {et~al.}(2015)\citenamefont {Qian},
  \citenamefont {Little}, \citenamefont {Howell},\ and\ \citenamefont
  {Eberly}}]{Qian:15}%
  \BibitemOpen
  \bibfield  {author} {\bibinfo {author} {\bibfnamefont {X.-F.}\ \bibnamefont
  {Qian}}, \bibinfo {author} {\bibfnamefont {B.}~\bibnamefont {Little}},
  \bibinfo {author} {\bibfnamefont {J.~C.}\ \bibnamefont {Howell}},\ and\
  \bibinfo {author} {\bibfnamefont {J.~H.}\ \bibnamefont {Eberly}},\ }\bibfield
   {title} {\bibinfo {title} {Shifting the quantum-classical boundary: theory
  and experiment for statistically classical optical fields},\ }\href@noop {}
  {\bibfield  {journal} {\bibinfo  {journal} {Optica}\ }\textbf {\bibinfo
  {volume} {2}},\ \bibinfo {pages} {611} (\bibinfo {year} {2015})}\BibitemShut
  {NoStop}%
\bibitem [{\citenamefont {Goldin}\ \emph {et~al.}(2010)\citenamefont {Goldin},
  \citenamefont {Francisco},\ and\ \citenamefont {Ledesma}}]{Goldin:10}%
  \BibitemOpen
  \bibfield  {author} {\bibinfo {author} {\bibfnamefont {M.~A.}\ \bibnamefont
  {Goldin}}, \bibinfo {author} {\bibfnamefont {D.}~\bibnamefont {Francisco}},\
  and\ \bibinfo {author} {\bibfnamefont {S.}~\bibnamefont {Ledesma}},\
  }\bibfield  {title} {\bibinfo {title} {Simulating bell inequality violations
  with classical optics encoded qubits},\ }\href@noop {} {\bibfield  {journal}
  {\bibinfo  {journal} {J. Opt. Soc. Am. B}\ }\textbf {\bibinfo {volume}
  {27}},\ \bibinfo {pages} {779} (\bibinfo {year} {2010})}\BibitemShut
  {NoStop}%
\bibitem [{\citenamefont {Karimi}\ and\ \citenamefont
  {Boyd}(2015)}]{Karimi1172}%
  \BibitemOpen
  \bibfield  {author} {\bibinfo {author} {\bibfnamefont {E.}~\bibnamefont
  {Karimi}}\ and\ \bibinfo {author} {\bibfnamefont {R.~W.}\ \bibnamefont
  {Boyd}},\ }\bibfield  {title} {\bibinfo {title} {Classical entanglement?},\
  }\href {https://doi.org/10.1126/science.aad7174} {\bibfield  {journal}
  {\bibinfo  {journal} {Science}\ }\textbf {\bibinfo {volume} {350}},\ \bibinfo
  {pages} {1172} (\bibinfo {year} {2015})}\BibitemShut {NoStop}%
\bibitem [{\citenamefont {T\"{o}ppel}\ \emph {et~al.}(2015)\citenamefont
  {T\"{o}ppel}, \citenamefont {Aiello}, \citenamefont {Marquardt},
  \citenamefont {Giacobino},\ and\ \citenamefont {Leuchs}}]{Toppel:15}%
  \BibitemOpen
  \bibfield  {author} {\bibinfo {author} {\bibfnamefont {F.}~\bibnamefont
  {T\"{o}ppel}}, \bibinfo {author} {\bibfnamefont {A.}~\bibnamefont {Aiello}},
  \bibinfo {author} {\bibfnamefont {C.}~\bibnamefont {Marquardt}}, \bibinfo
  {author} {\bibfnamefont {E.}~\bibnamefont {Giacobino}},\ and\ \bibinfo
  {author} {\bibfnamefont {G.}~\bibnamefont {Leuchs}},\ }\bibfield  {title}
  {\bibinfo {title} {Classical entanglement: Theory and application},\
  }\href@noop {} {\bibfield  {journal} {\bibinfo  {journal} {2015 European
  Conference on Lasers and Electro-Optics - European Quantum Electronics
  Conference}\ ,\ \bibinfo {pages} {EI3a3}} (\bibinfo {year}
  {2015})}\BibitemShut {NoStop}%
\bibitem [{\citenamefont {Spreeuw}(1998)}]{spre}%
  \BibitemOpen
  \bibfield  {author} {\bibinfo {author} {\bibfnamefont {R.}~\bibnamefont
  {Spreeuw}},\ }\bibfield  {title} {\bibinfo {title} {A classical analogy of
  entanglement},\ }\href@noop {} {\bibfield  {journal} {\bibinfo  {journal}
  {Foundations of Physics}\ }\textbf {\bibinfo {volume} {28}},\ \bibinfo
  {pages} {361} (\bibinfo {year} {1998})}\BibitemShut {NoStop}%
\bibitem [{\citenamefont {Khrennikov}(2020)}]{Khrennikov2020}%
  \BibitemOpen
  \bibfield  {author} {\bibinfo {author} {\bibfnamefont {A.}~\bibnamefont
  {Khrennikov}},\ }\bibfield  {title} {\bibinfo {title} {Quantum versus
  classical entanglement: Eliminating the issue of quantum nonlocality},\
  }\href {https://doi.org/10.1007/s10701-020-00319-7} {\bibfield  {journal}
  {\bibinfo  {journal} {Foundations of Physics}\ }\textbf {\bibinfo {volume}
  {50}},\ \bibinfo {pages} {1762} (\bibinfo {year} {2020})}\BibitemShut
  {NoStop}%
\bibitem [{\citenamefont {Glauber}(1963{\natexlab{b}})}]{PhysRev.131.2766}%
  \BibitemOpen
  \bibfield  {author} {\bibinfo {author} {\bibfnamefont {R.~J.}\ \bibnamefont
  {Glauber}},\ }\bibfield  {title} {\bibinfo {title} {Coherent and incoherent
  states of the radiation field},\ }\href@noop {} {\bibfield  {journal}
  {\bibinfo  {journal} {Phys. Rev.}\ }\textbf {\bibinfo {volume} {131}},\
  \bibinfo {pages} {2766} (\bibinfo {year} {1963}{\natexlab{b}})}\BibitemShut
  {NoStop}%
\bibitem [{\citenamefont {R{\v{e}}h{\'{a}}{\v{c}}ek}\ \emph
  {et~al.}(2010)\citenamefont {R{\v{e}}h{\'{a}}{\v{c}}ek}, \citenamefont
  {Mogilevtsev},\ and\ \citenamefont {Hradil}}]{PhysRevLett.105.010402}%
  \BibitemOpen
  \bibfield  {author} {\bibinfo {author} {\bibfnamefont {J.}~\bibnamefont
  {R{\v{e}}h{\'{a}}{\v{c}}ek}}, \bibinfo {author} {\bibfnamefont
  {D.}~\bibnamefont {Mogilevtsev}},\ and\ \bibinfo {author} {\bibfnamefont
  {Z.}~\bibnamefont {Hradil}},\ }\bibfield  {title} {\bibinfo {title}
  {Operational tomography: Fitting of data patterns},\ }\href@noop {}
  {\bibfield  {journal} {\bibinfo  {journal} {Phys. Rev. Lett.}\ }\textbf
  {\bibinfo {volume} {105}},\ \bibinfo {pages} {010402} (\bibinfo {year}
  {2010})}\BibitemShut {NoStop}%
\bibitem [{\citenamefont {Mogilevtsev}\ \emph {et~al.}(2013)\citenamefont
  {Mogilevtsev}, \citenamefont {Ignatenko}, \citenamefont {Maloshtan},
  \citenamefont {Stoklasa}, \citenamefont {Rehacek},\ and\ \citenamefont
  {Hradil}}]{Mogilevtsev_2013}%
  \BibitemOpen
  \bibfield  {author} {\bibinfo {author} {\bibfnamefont {D.}~\bibnamefont
  {Mogilevtsev}}, \bibinfo {author} {\bibfnamefont {A.}~\bibnamefont
  {Ignatenko}}, \bibinfo {author} {\bibfnamefont {A.}~\bibnamefont
  {Maloshtan}}, \bibinfo {author} {\bibfnamefont {B.}~\bibnamefont {Stoklasa}},
  \bibinfo {author} {\bibfnamefont {J.}~\bibnamefont {Rehacek}},\ and\ \bibinfo
  {author} {\bibfnamefont {Z.}~\bibnamefont {Hradil}},\ }\bibfield  {title}
  {\bibinfo {title} {Data pattern tomography: reconstruction with an unknown
  apparatus},\ }\href@noop {} {\bibfield  {journal} {\bibinfo  {journal} {New
  Journal of Physics}\ }\textbf {\bibinfo {volume} {15}},\ \bibinfo {pages}
  {025038} (\bibinfo {year} {2013})}\BibitemShut {NoStop}%
\bibitem [{\citenamefont {Mikhalychev}\ \emph {et~al.}(2015)\citenamefont
  {Mikhalychev}, \citenamefont {Mogilevtsev}, \citenamefont {Teo},
  \citenamefont {{\v R}eh{\'a}{\v c}ek},\ and\ \citenamefont
  {Hradil}}]{BayesianPRA2015}%
  \BibitemOpen
  \bibfield  {author} {\bibinfo {author} {\bibfnamefont {A.}~\bibnamefont
  {Mikhalychev}}, \bibinfo {author} {\bibfnamefont {D.}~\bibnamefont
  {Mogilevtsev}}, \bibinfo {author} {\bibfnamefont {Y.~S.}\ \bibnamefont
  {Teo}}, \bibinfo {author} {\bibfnamefont {J.}~\bibnamefont {{\v R}eh{\'a}{\v
  c}ek}},\ and\ \bibinfo {author} {\bibfnamefont {Z.}~\bibnamefont {Hradil}},\
  }\bibfield  {title} {\bibinfo {title} {Bayesian recursive data-pattern
  tomography},\ }\href {https://doi.org/10.1103/PhysRevA.92.052106} {\bibfield
  {journal} {\bibinfo  {journal} {Phys. Rev. A}\ }\textbf {\bibinfo {volume}
  {92}},\ \bibinfo {pages} {052106} (\bibinfo {year} {2015})}\BibitemShut
  {NoStop}%
\bibitem [{\citenamefont {Motka}\ \emph {et~al.}(2017)\citenamefont {Motka},
  \citenamefont {Pa{\'{u}}r}, \citenamefont {R{\v{e}}h{\'{a}}{\v{c}}ek},
  \citenamefont {Hradil},\ and\ \citenamefont
  {S{\'{a}}nchez-Soto}}]{Motka_2017}%
  \BibitemOpen
  \bibfield  {author} {\bibinfo {author} {\bibfnamefont {L.}~\bibnamefont
  {Motka}}, \bibinfo {author} {\bibfnamefont {M.}~\bibnamefont {Pa{\'{u}}r}},
  \bibinfo {author} {\bibfnamefont {J.}~\bibnamefont
  {R{\v{e}}h{\'{a}}{\v{c}}ek}}, \bibinfo {author} {\bibfnamefont
  {Z.}~\bibnamefont {Hradil}},\ and\ \bibinfo {author} {\bibfnamefont {L.~L.}\
  \bibnamefont {S{\'{a}}nchez-Soto}},\ }\bibfield  {title} {\bibinfo {title}
  {Efficient tomography with unknown detectors},\ }\href@noop {} {\bibfield
  {journal} {\bibinfo  {journal} {Quantum Science and Technology}\ }\textbf
  {\bibinfo {volume} {2}},\ \bibinfo {pages} {035003} (\bibinfo {year}
  {2017})}\BibitemShut {NoStop}%
\bibitem [{\citenamefont {Reut}\ \emph {et~al.}(2017)\citenamefont {Reut},
  \citenamefont {Mikhalychev},\ and\ \citenamefont {Mogilevtsev}}]{reut2017}%
  \BibitemOpen
  \bibfield  {author} {\bibinfo {author} {\bibfnamefont {V.}~\bibnamefont
  {Reut}}, \bibinfo {author} {\bibfnamefont {A.}~\bibnamefont {Mikhalychev}},\
  and\ \bibinfo {author} {\bibfnamefont {D.}~\bibnamefont {Mogilevtsev}},\
  }\bibfield  {title} {\bibinfo {title} {Data-pattern tomography of entangled
  states},\ }\href@noop {} {\bibfield  {journal} {\bibinfo  {journal} {Phys.
  Rev. A}\ }\textbf {\bibinfo {volume} {95}},\ \bibinfo {pages} {012123}
  (\bibinfo {year} {2017})}\BibitemShut {NoStop}%
\bibitem [{\citenamefont {Cooper}\ \emph {et~al.}(2014)\citenamefont {Cooper},
  \citenamefont {Karpiński},\ and\ \citenamefont {Smith}}]{pub.1039448817}%
  \BibitemOpen
  \bibfield  {author} {\bibinfo {author} {\bibfnamefont {M.}~\bibnamefont
  {Cooper}}, \bibinfo {author} {\bibfnamefont {M.}~\bibnamefont {Karpiński}},\
  and\ \bibinfo {author} {\bibfnamefont {B.~J.}\ \bibnamefont {Smith}},\
  }\bibfield  {title} {\bibinfo {title} {Local mapping of detector response for
  reliable quantum state estimation},\ }\href@noop {} {\bibfield  {journal}
  {\bibinfo  {journal} {Nature Communications}\ }\textbf {\bibinfo {volume}
  {5}},\ \bibinfo {pages} {4332} (\bibinfo {year} {2014})}\BibitemShut
  {NoStop}%
\bibitem [{\citenamefont {Slepyan}\ \emph {et~al.}(2021)\citenamefont
  {Slepyan}, \citenamefont {Vlasenko}, \citenamefont {Mogilevtsev},\ and\
  \citenamefont {Boag}}]{radars}%
  \BibitemOpen
  \bibfield  {author} {\bibinfo {author} {\bibfnamefont {G.}~\bibnamefont
  {Slepyan}}, \bibinfo {author} {\bibfnamefont {S.}~\bibnamefont {Vlasenko}},
  \bibinfo {author} {\bibfnamefont {D.}~\bibnamefont {Mogilevtsev}},\ and\
  \bibinfo {author} {\bibfnamefont {A.}~\bibnamefont {Boag}},\ }\bibfield
  {title} {\bibinfo {title} {Quantum radars and lidars: Concepts, realizations,
  and perspectives.},\ }\href {https://doi.org/10.1109/MAP.2021.3089994}
  {\bibfield  {journal} {\bibinfo  {journal} {IEEE Antennas and Propagation
  Magazine}\ ,\ \bibinfo {pages} {2}} (\bibinfo {year} {2021})}\BibitemShut
  {NoStop}%
\bibitem [{\citenamefont {Klauder}(1966)}]{PhysRevLett.16.534}%
  \BibitemOpen
  \bibfield  {author} {\bibinfo {author} {\bibfnamefont {J.~R.}\ \bibnamefont
  {Klauder}},\ }\bibfield  {title} {\bibinfo {title} {Improved version of
  optical equivalence theorem},\ }\href@noop {} {\bibfield  {journal} {\bibinfo
   {journal} {Phys. Rev. Lett.}\ }\textbf {\bibinfo {volume} {16}},\ \bibinfo
  {pages} {534} (\bibinfo {year} {1966})}\BibitemShut {NoStop}%
\bibitem [{\citenamefont {Lobino}\ \emph {et~al.}(2008)\citenamefont {Lobino},
  \citenamefont {Korystov}, \citenamefont {Kupchak}, \citenamefont {Figueroa},
  \citenamefont {Sanders},\ and\ \citenamefont {Lvovsky}}]{Lobino563}%
  \BibitemOpen
  \bibfield  {author} {\bibinfo {author} {\bibfnamefont {M.}~\bibnamefont
  {Lobino}}, \bibinfo {author} {\bibfnamefont {D.}~\bibnamefont {Korystov}},
  \bibinfo {author} {\bibfnamefont {C.}~\bibnamefont {Kupchak}}, \bibinfo
  {author} {\bibfnamefont {E.}~\bibnamefont {Figueroa}}, \bibinfo {author}
  {\bibfnamefont {B.~C.}\ \bibnamefont {Sanders}},\ and\ \bibinfo {author}
  {\bibfnamefont {A.~I.}\ \bibnamefont {Lvovsky}},\ }\bibfield  {title}
  {\bibinfo {title} {Complete characterization of quantum-optical processes},\
  }\href@noop {} {\bibfield  {journal} {\bibinfo  {journal} {Science}\ }\textbf
  {\bibinfo {volume} {322}},\ \bibinfo {pages} {563} (\bibinfo {year}
  {2008})}\BibitemShut {NoStop}%
\bibitem [{\citenamefont {Huelga}\ \emph {et~al.}(1997)\citenamefont {Huelga},
  \citenamefont {Macchiavello}, \citenamefont {Pellizzari}, \citenamefont
  {Ekert}, \citenamefont {Plenio},\ and\ \citenamefont
  {Cirac}}]{PhysRevLett.79.3865}%
  \BibitemOpen
  \bibfield  {author} {\bibinfo {author} {\bibfnamefont {S.~F.}\ \bibnamefont
  {Huelga}}, \bibinfo {author} {\bibfnamefont {C.}~\bibnamefont
  {Macchiavello}}, \bibinfo {author} {\bibfnamefont {T.}~\bibnamefont
  {Pellizzari}}, \bibinfo {author} {\bibfnamefont {A.~K.}\ \bibnamefont
  {Ekert}}, \bibinfo {author} {\bibfnamefont {M.~B.}\ \bibnamefont {Plenio}},\
  and\ \bibinfo {author} {\bibfnamefont {J.~I.}\ \bibnamefont {Cirac}},\
  }\bibfield  {title} {\bibinfo {title} {Improvement of frequency standards
  with quantum entanglement},\ }\href@noop {} {\bibfield  {journal} {\bibinfo
  {journal} {Phys. Rev. Lett.}\ }\textbf {\bibinfo {volume} {79}},\ \bibinfo
  {pages} {3865} (\bibinfo {year} {1997})}\BibitemShut {NoStop}%
\bibitem [{\citenamefont {Harder}\ \emph {et~al.}(2014)\citenamefont {Harder},
  \citenamefont {Mogilevtsev}, \citenamefont {Korolkova},\ and\ \citenamefont
  {Silberhorn}}]{PhysRevLett.113.070403}%
  \BibitemOpen
  \bibfield  {author} {\bibinfo {author} {\bibfnamefont {G.}~\bibnamefont
  {Harder}}, \bibinfo {author} {\bibfnamefont {D.}~\bibnamefont {Mogilevtsev}},
  \bibinfo {author} {\bibfnamefont {N.}~\bibnamefont {Korolkova}},\ and\
  \bibinfo {author} {\bibfnamefont {C.}~\bibnamefont {Silberhorn}},\ }\bibfield
   {title} {\bibinfo {title} {Tomography by noise},\ }\href@noop {} {\bibfield
  {journal} {\bibinfo  {journal} {Phys. Rev. Lett.}\ }\textbf {\bibinfo
  {volume} {113}},\ \bibinfo {pages} {070403} (\bibinfo {year}
  {2014})}\BibitemShut {NoStop}%
\bibitem [{\citenamefont {Tiedau}\ \emph {et~al.}(2018)\citenamefont {Tiedau},
  \citenamefont {Shchesnovich}, \citenamefont {Mogilevtsev}, \citenamefont
  {Ansari}, \citenamefont {Harder}, \citenamefont {Bartley}, \citenamefont
  {Korolkova},\ and\ \citenamefont {Silberhorn}}]{Tiedau_2018}%
  \BibitemOpen
  \bibfield  {author} {\bibinfo {author} {\bibfnamefont {J.}~\bibnamefont
  {Tiedau}}, \bibinfo {author} {\bibfnamefont {V.~S.}\ \bibnamefont
  {Shchesnovich}}, \bibinfo {author} {\bibfnamefont {D.}~\bibnamefont
  {Mogilevtsev}}, \bibinfo {author} {\bibfnamefont {V.}~\bibnamefont {Ansari}},
  \bibinfo {author} {\bibfnamefont {G.}~\bibnamefont {Harder}}, \bibinfo
  {author} {\bibfnamefont {T.~J.}\ \bibnamefont {Bartley}}, \bibinfo {author}
  {\bibfnamefont {N.}~\bibnamefont {Korolkova}},\ and\ \bibinfo {author}
  {\bibfnamefont {C.}~\bibnamefont {Silberhorn}},\ }\bibfield  {title}
  {\bibinfo {title} {Quantum state and mode profile tomography by the
  overlap},\ }\href@noop {} {\bibfield  {journal} {\bibinfo  {journal} {New
  Journal of Physics}\ }\textbf {\bibinfo {volume} {20}},\ \bibinfo {pages}
  {033003} (\bibinfo {year} {2018})}\BibitemShut {NoStop}%
\bibitem [{\citenamefont {Couteau}(2018)}]{spdc}%
  \BibitemOpen
  \bibfield  {author} {\bibinfo {author} {\bibfnamefont {C.}~\bibnamefont
  {Couteau}},\ }\bibfield  {title} {\bibinfo {title} {Spontaneous parametric
  down-conversion},\ }\href@noop {} {\bibfield  {journal} {\bibinfo  {journal}
  {Contemporary Physics}\ }\textbf {\bibinfo {volume} {59}},\ \bibinfo {pages}
  {291} (\bibinfo {year} {2018})}\BibitemShut {NoStop}%
\bibitem [{\citenamefont {Clauser}\ and\ \citenamefont
  {Horne}(1974)}]{PhysRevD.10.526}%
  \BibitemOpen
  \bibfield  {author} {\bibinfo {author} {\bibfnamefont {J.~F.}\ \bibnamefont
  {Clauser}}\ and\ \bibinfo {author} {\bibfnamefont {M.~A.}\ \bibnamefont
  {Horne}},\ }\bibfield  {title} {\bibinfo {title} {Experimental consequences
  of objective local theories},\ }\href@noop {} {\bibfield  {journal} {\bibinfo
   {journal} {Phys. Rev. D}\ }\textbf {\bibinfo {volume} {10}},\ \bibinfo
  {pages} {526} (\bibinfo {year} {1974})}\BibitemShut {NoStop}%
\bibitem [{\citenamefont {Wildfeuer}\ \emph {et~al.}(2007)\citenamefont
  {Wildfeuer}, \citenamefont {Lund},\ and\ \citenamefont
  {Dowling}}]{wildfeuer2007}%
  \BibitemOpen
  \bibfield  {author} {\bibinfo {author} {\bibfnamefont {C.~F.}\ \bibnamefont
  {Wildfeuer}}, \bibinfo {author} {\bibfnamefont {A.~P.}\ \bibnamefont
  {Lund}},\ and\ \bibinfo {author} {\bibfnamefont {J.~P.}\ \bibnamefont
  {Dowling}},\ }\bibfield  {title} {\bibinfo {title} {Strong violations of
  bell-type inequalities for path-entangled number states},\ }\href@noop {}
  {\bibfield  {journal} {\bibinfo  {journal} {Phys. Rev. A}\ }\textbf {\bibinfo
  {volume} {76}},\ \bibinfo {pages} {052101} (\bibinfo {year}
  {2007})}\BibitemShut {NoStop}%
\bibitem [{\citenamefont {Fuchs}\ and\ \citenamefont
  {Caves}(1995)}]{fuchs1995mathematical}%
  \BibitemOpen
  \bibfield  {author} {\bibinfo {author} {\bibfnamefont {C.~A.}\ \bibnamefont
  {Fuchs}}\ and\ \bibinfo {author} {\bibfnamefont {C.~M.}\ \bibnamefont
  {Caves}},\ }\bibfield  {title} {\bibinfo {title} {Mathematical techniques for
  quantum communication theory},\ }\href@noop {} {\bibfield  {journal}
  {\bibinfo  {journal} {Open Systems \& Information Dynamics}\ }\textbf
  {\bibinfo {volume} {3}},\ \bibinfo {pages} {345} (\bibinfo {year}
  {1995})}\BibitemShut {NoStop}%
\end{thebibliography}

%

 \end{document}